\documentclass[journal]{IEEEtran}
\pdfoutput=1
\usepackage[noadjust]{cite}
\usepackage{amsmath}
\usepackage{array}
\usepackage{mdwmath}
\usepackage{amssymb}
\usepackage{mathtools}
\usepackage{algorithm}
\usepackage[noend]{algpseudocode}
\usepackage{eqparbox}
\usepackage{stfloats}
\usepackage{url}
\usepackage[pdftex]{graphicx}
\graphicspath{{../pdf/}{../jpeg/}}
\DeclareGraphicsExtensions{.pdf,.jpeg,.png}
\usepackage{epstopdf}
\usepackage{amsfonts}

\usepackage{color}

\usepackage{tikz}
\usepackage{pgfplots}
\usepackage{subcaption}
\usetikzlibrary{spy}

\newtheorem{lemma}{Lemma}
\newtheorem{theorem}{Theorem}
\newtheorem{remark}{Remark}

\begin{document}

\title{\huge Intelligent Reflecting Surface Enabled Random Rotations Scheme for the MISO Broadcast Channel}
\author{Qurrat-Ul-Ain~Nadeem,~\IEEEmembership{Member,~IEEE,} Alessio Zappone,~\IEEEmembership{Senior Member,~IEEE,} and Anas~Chaaban,~\IEEEmembership{Senior Member,~IEEE}
        \thanks{This work was supported by the King Abdullah University of Science and Technology (KAUST) under Award OSR-2018-CRG7-3734.}
				\thanks{Q.-U.-A. Nadeem and A. Chaaban are with School of Engineering, University of British Columbia, Kelowna, Canada  (e-mail:  \{qurrat.nadeem, anas.chaaban\}@ubc.ca).}
\thanks{A. Zappone is with the University of Cassino and Southern Lazio, Cassino, Italy (alessio.zappone@unicas.it).}
}

\maketitle

\begin{abstract}

The current literature on intelligent reflecting surface (IRS) focuses on  optimizing the IRS phase shifts to yield coherent beamforming gains, under the assumption of perfect channel state information (CSI) of individual IRS-assisted links, which  is highly impractical.  This work, instead, considers the random rotations scheme at the IRS in which the reflecting elements only employ random phase rotations without requiring any CSI. The only CSI then needed is at the base station (BS) of the overall channel to implement the beamforming transmission scheme. Under this framework, we derive the sum-rate scaling laws in the large number of users regime for the IRS-assisted multiple-input single-output (MISO)  broadcast channel, with  optimal dirty paper coding (DPC) scheme and the lower-complexity random beamforming (RBF) and deterministic beamforming (DBF) schemes at the BS. The random rotations scheme increases the sum-rate by exploiting multi-user diversity, but also compromises the gain to some extent due to correlation. Finally, energy efficiency maximization problems in terms of the number of BS antennas, IRS elements and transmit power are solved using the derived scaling laws. Simulation results show the proposed scheme to improve the sum-rate, with performance becoming close to that under coherent beamforming for a large number of users.
\end{abstract}

\section{Introduction}
\label{Sec:Intro}

Although the emerging Fifth-Generation (5G) era has brought about a remarkable improvement in the speed, latency and reliability of cellular networks,  it still  faces challenges related to the high power consumption of the underlying technologies as well as the limited control over the wireless propagation environment \cite{6g_l}.   A promising new solution to address these issues is deploying intelligent reflecting surfaces (IRSs) \cite{SRE1, Renzo2019, aless, LIS} in existing wireless communication systems. The IRS is abstracted in the current literature as an array of nearly passive reflecting elements, where each element can  introduce a phase shift onto the incoming electromagnetic waves as directed by a smart controller  \cite{Renzo2019,aless,  SRE1,LIS, LIS_los, huang, annie, WSR, phy}. The resulting software-controlled reflections can tailor the propagation environment to yield desirable communication objectives, like increase the coverage, rates, energy efficiency etc, in a passive manner without generating new radio signals. 

Motivated by its numerous potential advantages, IRS-assisted communication has recently been investigated under different communication scenarios \cite{huang, LIS, annie, LIS_los, WSR, phy}. Several joint designs for precoding at the base station (BS) and reflect beamforming at the IRS have been proposed to achieve different communication goals, for example: maximize the system's energy efficiency subject to signal-to-interference-plus-noise ratio (SINR) constraints at the users in \cite{huang}, maximize the minimum user rate subject to a transmit (Tx) power constraint at the BS in \cite{annie}, minimize the Tx power subject to users'  SINR constraints in \cite{LIS}, maximize the sum-rate subject to a Tx power constraint in \cite{WSR}, and minimize the Tx power subject to a secrecy rate constraint in \cite{phy}.

All these works optimize the IRS phase-shifts assuming the availability of perfect channel state information (CSI) at the BS and IRS of the individual IRS-assisted links, that is the BS-IRS link and IRS-users links. However, reliable channel estimation of these individual links in IRS-assisted systems can cause a prohibitive training overhead due to the inability of the IRS to send, receive or process the pilots. The preliminary papers on channel estimation for these systems showed the required training time to estimate the individual IRS-assisted channels to grow  proportionally with the number of IRS elements \cite{LS, LS1, CE_MU, annie_OJ}, thus compromising the performance gains expected from deploying a large number of reflecting elements to a large extent.   Moreover, optimizing the IRS phase-shifts at the pace of fast-fading channel significantly increases the system complexity.

Motivated by these challenges, this paper presents an information-theoretic analysis of an IRS enabled  random rotations scheme in the multiple-input single-output (MISO) broadcast channel (BC). The proposed scheme requires the IRS elements to only introduce random phase rotations in each coherence interval without requiring instantaneous CSI. The BS will only require the CSI of the overall BS-users channels (or SINRs), instead of the individual IRS-assisted channels, to implement the precoding/beamforming scheme. The only other works that have analytically studied  the random rotations scheme at the IRS are \cite{rotate1} and \cite{dumb_me}, where the former develops low-complexity and energy-efficient transmission schemes for a  point-to-point IRS-assisted single-input single-output (SISO) system based on coding and selection approaches, while the latter studies the sum-rate of the SISO BC under opportunistic scheduling.

We define the MISO BC from an information-theoretic point-of-view as consisting of a single multi-antenna transmitter (or BS) communicating with multiple single-antenna users, where the BS transmits different data signals (intended for different users) from different transmitting antennas \cite{netit, RBF, DPC_ca, tareq, TDMA1}. For the MISO BC, the primary research focus has been on 1) quantifying the maximum achievable average sum-rate capacity, and 2) devising computationally efficient algorithms for capturing most of this capacity. The first question has been addressed using dirty paper coding (DPC) \cite{costa, DPC_ca}, which solves the MISO BC problem optimally. The average sum-rate capacity (referred to as sum-capacity in this paper) achieved by DPC in a Gaussian BC is shown to scale as $M \log \log K$, with the number of users $K$ for a fixed number of Tx antennas $M$ \cite{TDMA1, RBF}.  While the capacity  increase is linear in $M$, DPC is  computationally expensive and requires full CSI, motivating the development of sub-optimal schemes that use partial CSI.

A popular scheme that addresses these challenges is random beamforming (RBF), which constructs $M$ random orthonormal beams and on each beam transmits to the user with the highest SINR.  At the start of a coherence interval, each user measures $M$ downlink SINR values corresponding to the pilot symbols transmitted on the $M$ beams by the BS,  and feeds back only one real number (its best SINR) and the corresponding beam index. The feedback overhead is therefore much less than sending back $M$ complex numbers associated with the full channel information  required by DPC. The BS then schedules the best user to transmit to on each beam.   Interestingly, the sum average rate scaling of RBF is shown to asymptotically coincide with the average sum-capacity scaling achieved by DPC, i.e.  the sum average rate achieved by RBF also scales as $M \log \log K$ \cite{RBF}. The gain with $K$ is explained by the multi-user (MU)-diversity effect \cite{dumb}, i.e. in a system with many users with independently time-varying channels, it is very likely to have at each time some users whose SINRs are much higher than the average SINR and by scheduling these users, the sum-rate can be significantly increased. This MU-diversity effect can be enhanced by increasing the dynamic range of channel fluctuations. We propose to do this using the time-varying random rotations introduced by an IRS into the BS-users channels. Our idea is inspired from the rotate-and-forward protocol proposed in \cite{rotate} for the slow-fading relay channel, where artificial fast fading created using the random rotations at the relay  resulted in  an optimal diversity-multiplexing tradeoff. 

Motivated by these works, we consider incorporating an IRS into the conventional MISO BC, where the BS-IRS channel is line-of-sight (LoS), the IRS-users channels and BS-users direct channels are Rayleigh faded and the IRS elements induce only random phase rotations. The assumption that the BS-IRS channel is LoS dominated has also been made in many other works such as \cite{LIS_los, LoS1, LoS2, LoS4, LoS5, LoS6, annie, annie_OJ, CE_MU} and is quite practical given the BS and IRS have fixed positions with few obstacles around. The overriding question then is to study the effect of the IRS enabled random rotations scheme on the average sum-capacity scaling (achieved by DPC) and the sum average rate scaling achieved by RBF. We carry out an asymptotic study of the sum-rate in the limit of a large number of users, leveraging results from extreme value theory, and find that the random phase rotations at the IRS increase the MU-diversity gain. At the same time, the spatial correlation introduced  in the overall channel by the BS-IRS LoS link reduces the sum-rate \cite{tareq}. We study the interplay between the MU-diversity gain and the rate loss due to correlation  in the developed scaling expressions. Under a deterministic variation of RBF, referred to as deterministic beamforming (DBF), the sum average rate scaling is shown to coincide with the average sum-capacity scaling.  Simulation results show  a significant sum-rate gain with the random rotations scheme. Interestingly, the performance gap between the proposed scheme and coherent beamforming (under which the IRS phase shifts are designed based on full CSI) is shown to significantly reduce for large $K$. 

We also  optimize the system in terms of energy efficiency (EE) by obtaining the optimal system configuration parameters, including the number of BS antennas, the number of IRS elements and the total Tx power, that maximize the EE. The developed sum average rate scaling expression for DBF along with a realistic power consumption model are used to formulate the EE scaling. The challenging problem, that involves two discrete variables appearing as upper limits of sums and products, is solved using alternating optimization and line search methods subject to constraints on the maximum values of the parameters.  Simulation results show the EE performance of the IRS-assisted system to be significantly better than the conventional system. 

The paper is organized as follows. In Sec. \ref{Sec:Sys}, the MISO BC and IRS enabled random rotations scheme are outlined. In Sec. \ref{Sec:Asym}, the average sum-capacity scaling law for DPC and the sum average rate scaling laws for RBF and DBF are derived. In Sec. \ref{Sec:Opt} we maximize the EE with respect to the number of BS antennas, IRS elements and Tx power. Simulation results and conclusions are provided in Sec. \ref{Sec:Sim} and Sec. \ref{Sec:Con} respectively.

\section{System Model}
\label{Sec:Sys}
In this section, we outline the transmission model for the IRS-assisted MISO BC.

\subsection{MISO BC}
\label{Sec:Sys1}

Consider a Gaussian BC from an $M$-antenna BS to $K$ single-antenna users under the block-fading model, where the channel is constant during a coherence interval of length $T$ time-slots and varies independently from one such interval to the next.  Let $\mathbf{s}(t)$ be the $M\times 1$ vector of Tx symbols in time-slot $t$, then the received signal, $y_k(t)\in \mathbb{C}$, at user $k$ in time-slot $t$ is given as
\begin{align}
\label{rx}
y_k(t)=\mathbf{h}_k^H \mathbf{s}(t) + n_k(t), \hspace{.1in} t=1,\dots, T,
\end{align}
where  $\mathbf{h}_{k} \in \mathbb{C}^{M\times 1}$ is the channel from the BS to user $k$ and $n_k(t)$ is the complex Gaussian noise with zero mean and variance $\sigma^2$. We assume that the average Tx power, denoted as $P_T$, is fixed at at all times and therefore the Tx signal vector $\mathbf{s}(t)$ must satisfy the power constraint $\mathbb{E}[\mathbf{s}(t)^H \mathbf{s}(t)]=P_T$, $\forall t$. Denoting the average rate of user $k$ as $R_k$, where the average is taken over the fading distribution of $(\mathbf{h}_1, \dots, \mathbf{h}_K)$, we are interested in analyzing the behavior of the downlink sum average rate, i.e., $\sum_{k=1}^K R_k$.

\vspace{-.2in}
\subsection{IRS-Assisted Channel Model}
\label{Sec:Sys2}

\begin{figure}
\centering
\includegraphics[scale=.25]{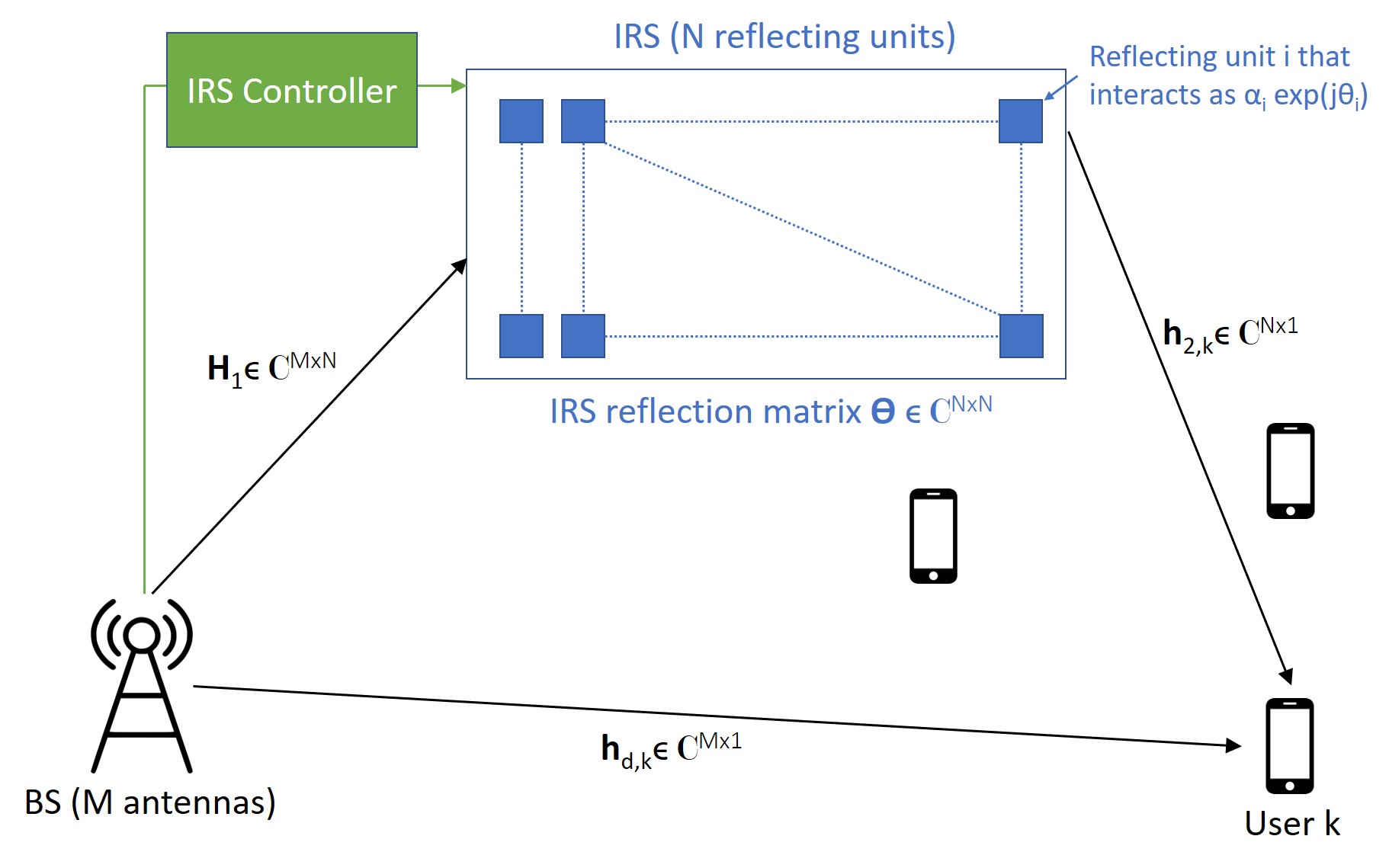}
\caption{IRS-assisted MISO BC.}
\label{LIS_model}
\end{figure}

The communication from the BS to users is assisted by an IRS, composed of $N$ passive reflecting elements (See Fig. \ref{LIS_model}).  The resulting $M\times 1$ channel between the BS and user $k$ is 
\begin{align}
\label{ch_RS_RB}
&\mathbf{h}_{k}=\sqrt{\beta_{r,k}}\mathbf{H}_1 \boldsymbol{\Theta} \mathbf{h}_{2,k}+\sqrt{\beta_{d,k}} \mathbf{h}_{d,k},
\end{align}
where  $\mathbf{H}_{1} \in \mathbb{C}^{M\times N}$ is the channel between the BS and the IRS, $\mathbf{h}_{2,k} \in \mathbb{C}^{N\times 1}$ is the channel  between the IRS and user $k$, $\mathbf{h}_{d,k}\in \mathbb{C}^{M\times 1}$ is the direct channel  between the BS and user $k$, and $\boldsymbol{\Theta}= \text{diag}(\alpha_1 \exp(j\theta_{1}), \dots, \alpha_N \exp(j\theta_{N})) \in \mathbb{C}^{N\times N}$ accounts for the IRS response. Here $\theta_{n}\in [0, 2\pi]$ is the phase shift applied by element $n$ and $\alpha_n \in [0,1]$ is the reflection coefficient of element $n$ which is fixed and depends on the IRS construction. Under the random rotations scheme, each $\theta_n$ is  randomly drawn  from the uniform distribution over the interval $[0,2\pi]$, with a new set of phase-shifts applied in each coherence interval.   Moreover $\beta_{r,k}=\beta_1 \beta_{2,k}$, where $\beta_1$, $\beta_{2,k}$ and $\beta_{d,k}$ are the path loss factors for BS-IRS, IRS-user-$k$ and direct links respectively. 

The channel between the BS and IRS is considered to be LoS dominated. This assumption, made in many other related works \cite{LIS_los, LoS1, LoS2, LoS4, LoS5, LoS6, annie, annie_OJ}, is supported in literature using two points. First, the LoS path between the BS and the IRS will usually always exist. The BS tower is generally elevated high and the IRS is also envisioned to be integrated onto tall structures in the environment, so both will have a few obstacles around. Given the positions of BS and IRS are fixed, a stable LoS channel between the BS and the IRS is expected to exist and can be constructed  using the directional angular information.  Second, the path loss for NLoS paths is much larger than that for the LoS path in the next generation systems. In fact it is noted that in mmWave systems, the typical value of Rician factor (ratio of energy in LoS to that in NLoS component) is between $20$\rm{dB} and $40$\rm{dB} \cite{LoS1}, which is sufficiently large to neglect any NLoS channel components in $\mathbf{H}_1$ as compared to the LoS component. Under these remarks, we assume that the BS-IRS channel is LoS dominated and neglect any NLoS paths in this channel. 

A uniform rectangular array (URA) of $N=N_1\times N_2$ reflecting elements is considered at the IRS, where $N_1$ and $N_2$ are the number of elements placed with inter-element spacing $d_{IRS}^{(1)}$ and $d_{IRS}^{(2)}$ along the two principal directions of the URA, characterized by unit vectors $\hat{\mathbf{n}}_1(\phi_{IRS}, \theta_{IRS})$ and $\hat{\mathbf{n}}_2(\phi_{IRS}, \theta_{IRS})$ respectively. Here ($\phi_{IRS}, \theta_{IRS}$) are the LoS azimuth and elevation angles that describe these principal directions at the IRS. A uniform linear array (ULA) is considered at the BS, with antennas placed at a spacing of $d_{BS}$ units along $\hat{\mathbf{n}}_1(\phi_{BS}, \theta_{BS})$, described by the LoS angles ($\phi_{BS}, \theta_{BS}$). Under the spherical wave model, the entries of the LoS channel $\mathbf{H}_1$ are computed as \cite{losura}
\begin{align}
\label{H_1}
&[\mathbf{H}_1]_{m,n}=\exp\left(j \frac{2\pi}{\lambda} l_{(m),(n_1,n_2)}\right), \hspace{.03in} m=1,\dots, M, \\
& n=1,\dots, N, n=n_1N_2+n_2+1, \nonumber
\end{align}
where $n_1=0,\dots, N_1-1$, $n_2=0, \dots, N_2-1$,  and $l_{(m),(n_1,n_2)}$ is the path length between the BS antenna $m$ and IRS element $(n_1,n_2)$, given as $l_{(m),(n_1,n_2)}=||\mathbf{a}_{IRS}^{(n_1,n_2)}-\mathbf{a}_{BS}^{(m)}||$. Here $\mathbf{a}_{IRS}^{(n_1,n_2)}$ is the steering vector from the global origin (defined as first antenna in the BS ULA) to the IRS element $(n_1, n_2)$, and $\mathbf{a}_{BS}^{(m)}$ is the steering vector from global origin to the BS antenna $m$. The expressions of $\mathbf{a}_{BS}^{(m)}$ and $\mathbf{a}_{IRS}^{(n_1,n_2)}$  can be found in \cite[equations (12) and (13)]{losura} and depend on $d_{IRS}^{(1)}$, $d_{IRS}^{(2)}$, $d_{BS}$, the distance $\bar{D}$ between the BS and IRS as well as the expressions of the unit vectors $\hat{\mathbf{n}}_1(\phi, \theta)$ and $\hat{\mathbf{n}}_2(\phi, \theta)$ provided in \cite[equations (6) and (9)]{losura}.  An important observation is that we have taken the spherical nature of the electromagnetic wave propagation into account, by applying the actual distance between the Tx antennas and receive (Rx) elements when considering the received LoS phase in \eqref{H_1}. Consequently, there is no restriction on the rank of $\mathbf{H}_1$, which can generally be of high rank for moderate BS-IRS distances and large $N$ \cite{losura}.

Also, we consider the IRS-users channels $\mathbf{h}_{2,k}$'s and BS-users direct channels $\mathbf{h}_{d,k}$'s to undergo Rayleigh fading. This is justified by observing that the LoS paths in these channels are usually blocked due to the scattering and blocking structures/objects around the ground users. These links are therefore characterized by NLoS paths and are modeled as $\mathbf{h}_{2,k} \sim \mathcal{CN}(\mathbf{0},\mathbf{I}_{N})$ and $\mathbf{h}_{d,k} \sim \mathcal{CN}(\mathbf{0},\mathbf{I}_{M})$, with the channels being independent across the users. The overall channel $\mathbf{h}_k$ in \eqref{ch_RS_RB} is a sum of two zero-mean complex Gaussian vectors, where the second vector (i.e. $\sqrt{\beta}_{d,k}\mathbf{h}_{d,k}$) has uncorrelated elements. The first vector has correlated entries with the correlation matrix found by conditioning the expectation on $\boldsymbol{\Theta}$ first as $\mathbb{E}[\beta_{r,k} \mathbf{H}_1 \boldsymbol{\Theta} \mathbf{h}_{2,k}\mathbf{h}_{2,k}^H \boldsymbol{\Theta}^H \mathbf{H}_1^H ]\hspace{-.05in}=\beta_{r,k} \mathbf{H}_1 \mathbb{E}_{\boldsymbol{\Theta}}[\mathbb{E}_{\mathbf{h}_{2,k}|\boldsymbol{\Theta}}[ \boldsymbol{\Theta} \mathbf{h}_{2,k}\mathbf{h}_{2,k}^H \boldsymbol{\Theta}^H |\boldsymbol{\Theta}]]\mathbf{H}_1^H \hspace{-.07in}=\hspace{-.07in}\beta_{r,k} \mathbf{H}_1 \mathbb{E}_{\boldsymbol{\Theta}}[ \boldsymbol{\Theta} \mathbf{I}_N \boldsymbol{\Theta}^H ]\mathbf{H}_1^H\hspace{-.09in}=\hspace{-.07in}\beta_{r,k}\mathbf{H}_1 \text{diag}(\alpha_1^2 ,\dots, \alpha_N^2) \mathbf{H}_1^H$. Therefore, the overall channel in (\ref{ch_RS_RB}) under the random rotations scheme is distributed as \vspace{-.1in}
\begin{align}
\label{ch_corr_RBF}
&\mathbf{h}_{k} \sim \mathcal{CN}(\mathbf{0},\mathbf{R}_k),
\end{align}
where $\mathbf{R}_k=\mathbb{E}[\mathbf{h}_k \mathbf{h}_k^H]=\beta_{r,k}\mathbf{H}_1  \text{diag}(\alpha_1^2 ,\dots, \alpha_N^2) \mathbf{H}_1^H+\beta_{d,k} \mathbf{I}_M \in \mathbb{C}^{M\times M}$ is the covariance matrix of the overall channel. It can be written using \eqref{H_1} as \vspace{-.07in}
\begin{align}
\label{R_def}
&[\mathbf{R}_k]_{m,m'}\hspace{-.04in}=\hspace{-.04in}\begin{cases} \beta_{r,k} \sum_{n=1}^N \alpha_n^2+\beta_{d,k}, & \hspace{-.06in} \text{if } m=m', \\
\beta_{r,k}  \sum_{n=1}^N \alpha_n^2 \exp\Big(j \frac{2\pi}{\lambda} (l_{(m),(n_1,n_2)}\\ -\l_{(m'),(n_1,n_2)})\Big), & \hspace{-.06in} \text{if } m\neq m'. \end{cases}
\end{align}

\section{Asymptotic Analysis of the Sum-Rate}
\label{Sec:Asym}

We now study the scaling laws of the average sum-capacity (achieved by DPC) and the sum average rate achieved by RBF and DBF against the number of users $K$, under the proposed framework using the following result from extreme value theory.

\begin{lemma}\label{Lemma1} \cite[Lemma 2]{dumb}: Let $x_{1},\dots,x_{K}$ be independent and identically distributed (i.i.d.) random variables (RVs) with a common cdf $F_{X}(x)$ and pdf $f_{X}(x)$, satisfying $F_{X}(x)<1$ for all finite $x$ and is twice differentiable for all $x$. The distributions are such that they satisfy
\begin{align}
\label{growth}
&\underset{x\rightarrow \infty}{\text{lim}}  g(x) =c>0,
\end{align}
for some constant $c$, where $g(x)=\frac{1-F_{X}(x)}{f_{X}(x)}$. Then for $l_k$ satisfying $F(l_k)=1-1/K$, we have \vspace{-.1in}
\begin{align}
&\underset{k=1,\dots,K}{\text{max}}x_{k}-l_{K}, 
\end{align}
 converges in distribution to a limiting random variable $Z$ with CDF $F_{Z}(z)=\exp(-\exp(-z/c))$. 
\end{lemma}
This lemma states that the maximum of $K$ i.i.d. RVs described above grows like $l_K$ as $K\rightarrow \infty$.

 To study the asymptotic behaviour of the sum-rate, we will assume a homogeneous network  as done in many works including \cite{RBF, tareq, TDMA1, dumb, usergroups}. The assumption is stated below.

\textit{Assumption 1:} We assume $\beta_{r,k}=\beta_r$, $\beta_{d,k}=\beta_d$, $\forall k$ resulting in $\mathbf{R}_k=\mathbf{R}$, $\forall k$. 

Note that the users' channels in \eqref{ch_corr_RBF} are independent by noting that $\mathbf{E}[\mathbf{h}_{k} \mathbf{h}^H_{k'}]=\mathbf{0}$ for $k\neq k'$. \textit{Assumption 1} will ensure that they are identically distributed as well. This makes the analysis based on extreme value theory tools tractable since \textit{Lemma 1} is applicable to i.i.d. RVs.

\vspace{-.1in}
\subsection{Average Sum-Capacity Scaling of the Proposed Framework} \vspace{-.05in}
\label{Sec:Asym1}
For the case where full CSI is available at the BS and users, it has been shown that the average sum-capacity of the Gaussian BC can be achieved using DPC. Intuitively, if the BS knows the channels $\mathbf{h}_k$'s of all users, it can use DPC to code against the interference for each user while preserving the power constraint. The average sum-capacity $R_{sum}$ is written as \cite{RBF} \vspace{-.1in}
\begin{align}
\label{rdpc}
&R_{sum}=\mathbb{E}\left[\underset{\underset{\sum_{k=1}^K p_k = P_T}{p_1, \dots, p_K}}{\text{max}} \log \text{det} \left( \mathbf{I}+\frac{1}{\sigma^2}\sum_{k=1}^K   p_k \mathbf{h}_k \mathbf{h}_k^H \right)  \right],
\end{align}
where $p_1, \dots, p_K$ is the optimal power allocation and the expectation is performed with respect to the distribution of $\mathbf{h}_k$'s. In a system without IRS, $R_{sum}$ is shown to scale for large $K$  as \cite{RBF}
\begin{align}
\label{DPC_sc}
&R^{(K)}_{sum}=M \log (P_T/(\sigma^2 M))+M \log (\beta_d \log K)+ o(1).
\end{align}

Here, we develop the average sum-capacity scaling law for the  IRS-assisted MISO BC under the random rotations scheme. Intuitively, different from \cite{RBF}, the sum-capacity will be affected 1) positively by the IRS array gain as highlighted through the sum over $N$ in the diagonal elements of \eqref{R_def}, and 2) adversely by the correlation in the IRS-assisted channel in \eqref{ch_corr_RBF}. The proof of the scaling law (presented in the following theorem) starts by establishing  an upper bound on the average sum-capacity scaling and showing that this bound is achievable using a low-complexity  DBF scheme outlined in Sec. \ref{Sec:Asym3}.

\begin{theorem} \label{Thm1} Consider the Gaussian BC given in \eqref{ch_RS_RB} comprising of an $M$-antenna BS  serving $K$ single-antenna users. This communication is assisted by an $N$-element IRS inducing only random phase rotations. Assume that the BS and users have perfect CSI of $\mathbf{h}_k$'s. Let $M$, $N$ and $\sigma^2$ be fixed, then for large $K$ and under \textit{Assumption 1}, the average sum-capacity scales as\vspace{-.1in}
\begin{align}
\label{DPC_sc_RS}
&R^{(K)}_{sum}=M \log \left(\frac{P_T }{\sigma^2 M}\right)+ M\log ((\beta_r\sum_{n=1}^N \alpha_n^2 +\beta_d) \log K) \nonumber \\
&+ M \log (\text{det } \bar{\mathbf{R}})^{1/M}+ o(1),
\end{align}
where $\bar{\mathbf{R}}=\frac{1}{\beta_r \sum_{n=1}^N \alpha_n^2 +\beta_d}\mathbf{R}$.
\end{theorem}
\begin{IEEEproof}
The proof is provided in Appendix \ref{App:DPC}.
\end{IEEEproof}

We observe that $\log (\text{det } \bar{\mathbf{R}}) \leq 0$, with equality when $\bar{\mathbf{R}}$ is an identity matrix. This is because the sum of the eigenvalues of $\bar{\mathbf{R}}$ is $M$ since $\text{tr}(\bar{\mathbf{R}})=M$ (Observe using \eqref{R_def} that all diagonal elements of $\bar{\mathbf{R}}$ will equal 1). Therefore, the product of these eigenvalues, which is $\text{det } \bar{\mathbf{R}}$, would be less than or equal to $1$. As compared to the result in (\ref{DPC_sc}) for a MISO BC without IRS, the average sum-capacity is increased by $M\log (\frac{\beta_r }{\beta_d}\sum_{n=1}^N \alpha_n^2+1)$ and decreased by a constant $|\log (\text{det } \bar{\mathbf{R}})|$, when an IRS introducing random phase rotations is incorporated into the system.  However as $N$ increases, one can expect the positive effect of $M\log (\frac{\beta_r}{\beta_d}\sum_{n=1}^N\alpha_n^2+1)$ to dominate, resulting in significant performance gains by deploying an IRS in the MISO BC.
 
There are two major drawbacks of DPC. First, it is computationally very complex, both at the BS and users. Moreover, it requires full CSI feedback of the overall channel $\mathbf{h}_k \in \mathbb{C}^{M\times 1}$ from each user to the BS resulting in prohibitively high feedback overhead when $M$ and $K$ are large. Therefore, research has focused on devising schemes that impose less computational complexity and feedback requirements but still achieve most of the sum-rate promised by DPC. The computational cost and feedback overhead associated with DPC are discussed in Sec. III-D.

\begin{remark} When each user has $L$ receive antennas, the average sum-capacity (of DPC) scales  as $R^{(K)}_{sum}=M \log \left(\frac{P_T }{\sigma^2 M}\right)+ M\log ((\beta_r\sum_{n=1}^N \alpha_n^2 +\beta_d) \log (L K)) + M \log (\text{det } \bar{\mathbf{R}})^{1/M}+ o(1)$. The result can be proved using a similar extension as done in \cite{TDMA1}.
\end{remark}

\vspace{-.15in}
\subsection{Random Beamforming}
\label{Sec:Asym2}
\vspace{-.05in}
To serve multiple users simultaneously without having full CSI at the BS, the RBF scheme was proposed in \cite{RBF} that constructs $M$ random beams and on each beam transmits to the user with the highest SINR. The only feedback required from each user is its maximum SINR  along with the beam index on which the SINR is maximized, instead of $M$ complex numbers representing the estimate of $\mathbf{h}_k$ as required by DPC. The process of RBF is explained in detail below.

At the start of each coherence interval,  the BS generates $M$ random orthonormal beamforming vectors $\boldsymbol{\phi}_{m}$ of size $M\times 1$ for $m=1,\dots, M$,  according to an isotropic distribution.  A straightforward way to generate the  RBF matrix, $\boldsymbol{\Phi}=[\boldsymbol{\phi}_{1}, \dots, \boldsymbol{\phi}_{M}]$, containing the  $M$ isotropically distributed (i.d.) orthonormal beamforming vectors is outlined in \cite{isot}. The method entails that we first generate a $M\times M$ random matrix $\mathbf{Y}$ whose elements are i.i.d. $\mathcal{CN}(0,1)$ RVs and then perform its QR (Gram-Schmidt) factorization as $\mathbf{Y}=\boldsymbol{\Phi} \mathbf{T}$, where $\boldsymbol{\Phi}$ is an i.d. unitary matrix and $\mathbf{T}$ is an upper triangular matrix. This yields the RBF matrix $\boldsymbol{\Phi}$ as $\boldsymbol{\Phi}=\mathbf{Y}\mathbf{T}^{-1}$. The proof that $\boldsymbol{\Phi}$ is indeed i.d. with orthonormal vectors can be found in \cite[Appendix A]{isot}.

Then at time-slot $t$, the BS multiplies the $m^{th}$ beamforming vector $\boldsymbol{\phi}_m$ with the $m^{th}$ transmit data symbol denoted as $\tilde{s}_m(t)$, such that the transmitted signal vector is given as
\begin{align}
\label{sig}
&\mathbf{s}(t)=\sum_{m=1}^M \boldsymbol{\phi}_{m}(t) \tilde{s}_m(t), \hspace{.1in} t=1,\dots, T.
\end{align}
In this paper,  we assume the data symbols  to be i.i.d. letters from codewords of a Gaussian capacity-achieving codebook with $\mathbb{E}[\tilde{\mathbf{s}}(t)\tilde{\mathbf{s}}(t)^H]=\frac{P_T}{M}\mathbf{I}_M$, where $\tilde{\mathbf{s}}(t)=[\tilde{s}_1(t), \dots, \tilde{s}_M(t)]^T$. The Tx signal vector then satisfies the power constraint $\mathbb{E}[\mathbf{s}(t)^H \mathbf{s}(t)]=P_T$.  After the coherence interval of $T$ channel uses, the BS chooses another set of orthonormal vectors $\{\boldsymbol{\phi}_m \}$ and constructs the signal in \eqref{sig} and so on. From now on, we drop the time index $t$ for simplicity. 

The RBF scheme allows $M$ users to be served simultaneously on the $M$ random beams (i.e. on the $M$ beamforming vectors $\boldsymbol{\phi}_m$'s). These $M$ out of $K$ users are selected based on their SINR feedback. Specifically during the training phase at the start of a coherence interval, the BS transmits orthogonal pilot symbols on the $M$ random beams $\boldsymbol{\phi}_m$, $m=1,\dots, M$. Denoting the pilot symbol transmitted on beam $m$ as $\tilde{s}^{(p)}_m$, with the pilot symbols satisfying the same power constraint as  data symbols, the received signal $y_{k}$ at user $k$ during the training phase is 
\begin{align}
y_k=\sum_{m=1}^M \mathbf{h}_{k}^H \boldsymbol{\phi}_{m} \tilde{s}^{(p)}_m + n_k.
\end{align}
User $k$ then computes the following $M$ SINRs by assuming that $\tilde{s}^{(p)}_m$ is the desired signal and the others are interfering signals:
\begin{align}
\label{SINR_RBF}
&\gamma_{k,m}=\frac{|\mathbf{h}_{k}^H \boldsymbol{\phi}_{m}|^2}{\frac{M \sigma^2}{P_T} + \sum_{i\neq m}|\mathbf{h}_{k}^H \boldsymbol{\phi}_{i}|^2}, \hspace{.1in} m=1,\dots, M.
\end{align}
Note that the SINRs above, corresponding to $m=1,\dots, M$ beams, are computed at the user by correlating the received signal $y_k$ in (12) one-by-one with each pilot symbol $\tilde{s}^{(p)}_m$ as $y_k \tilde{s}^{(p)^*}_m$. Given the orthogonality of pilot symbols, this allows the users to estimate $\mathbf{h}_{k}^H \boldsymbol{\phi}_{m}$ corresponding to $m=1,\dots, M$ beams, and hence compute the $\gamma_{k,m}$'s defined in \eqref{SINR_RBF}. Also note that the beams are associated with pilot symbols. Therefore, the user uses the pilot symbols to know which beam (i.e. which $\boldsymbol{\phi}_m$) it is processing and get the corresponding SINR for that beam.

Once the user has computed $\gamma_{k,m}$ for $m=1,\dots, M$ beams, it feeds back its maximum SINR, i.e. $\underset{m=1,\dots, M}{\text{max}} \gamma_{k,m}$, along with the beam index $m$ (i.e. the index of  $\boldsymbol{\phi}_m$) on which the SINR is maximized. Once the BS has this partial CSI from all  users, it assigns the beamforming vector $\boldsymbol{\phi}_m$ to transmit data symbol $\tilde{s}_m$ to the user with the highest corresponding SINR, i.e. on beamforming vector $\boldsymbol{\phi}_m$ serve the user $k'=\underset{k=1,\dots, K}{\text{arg max}} \gamma_{k,m}$. Using this scheduling, the BS constructs the signal vector in \eqref{sig}. The sum average rate during downlink transmission can then be written as \cite{RBF}\footnote{More rigorously the maximization in \eqref{exp} over $K$ should be performed over the subset (or fraction $\epsilon_m$) of users who report $m$ as the maximizing beam index. In practice, the users will not report beam index $m$ when either this was not the beam on which they observed the maximum SINR or they did not receive that beam (due to channel being orthogonal to $\boldsymbol{\phi}_m$). However as $K$ grows large, the impact of this fraction  on the average rate scaling achieved on each beam will grow to zero because $\log (\epsilon_m K)\approx \log K$ as $K\rightarrow \infty$, which is why the definition in \eqref{exp} with $o(1)$ term added is utilized \cite{RBF, tareq, usergroups, TDMA1}.} \vspace{-.1in}
\begin{align}
\label{exp}
&R_{RBF}=\sum_{m=1}^M \mathbb{E}\left[\log \left(1+ \underset{k=1,\dots, K}{\text{max}} \gamma_{k,m} \right)  \right]+o(1),
\end{align}
 where $\gamma_{k,m}$ is given by \eqref{SINR_RBF} and $o(1)$ represents the terms that go to zero as $K\rightarrow \infty$.

Note that if user $k$ has maximum SINR on two beams then the BS has to schedule another (weaker) user on one of these two beams resulting in a decrease in the sum average rate. However, it is shown in \cite{RBF} that  the probability that user $k$ has maximum SINR on two beams goes to zero as $K$ grows large, and therefore the sum average rate is given by \eqref{exp} with $o(1)$ added to represent the terms that go to zero with $K$. 

The authors in \cite{RBF} studied the behaviour of (\ref{exp}) for a MISO Gaussian BC without an IRS, where $\mathbf{h}_k\sim \mathcal{CN}(\mathbf{0},\beta_d \mathbf{I}_M)$, and showed the resulting scaling law with $K$ to be given as
\begin{align}
\label{RBF_sc}
&R^{(K)}_{RBF}=M \log (\beta_d \log K) + M \log \frac{P_T}{M \sigma^2}+ o(1).
\end{align}

In this section, we will investigate the impact of the random rotations IRS scheme on this scaling law. In \cite{RBF}, the sum average rate does not depend on the distribution of $\boldsymbol{\phi}_m$'s as multiplying the circularly symmetric complex Gaussian vector $\mathbf{h}_k\sim \mathcal{CN}(\mathbf{0},\beta_d \mathbf{I}_M)$ with a unitary vector does not change its distribution. However for our correlated scenario, the expectation in (\ref{exp}) depends on both the distribution of $\boldsymbol{\phi}_m$'s and the distribution of $\mathbf{h}_k$'s and is written as \vspace{-.1in}
\begin{align}
\label{exp1}
R_{RBF}&=\sum_{m=1}^M \mathbb{E}_{\boldsymbol{\Phi}}\left[ \mathbb{E}_{\mathbf{h}_{k}'s|\boldsymbol{\Phi}}\left[\log \left(1+ \underset{k=1,\dots, K}{\text{max}} \gamma_{k,m} \right) \right]|\boldsymbol{\Phi} \right]\nonumber \\
&+o(1),
\end{align}
 i.e., we first condition on $\boldsymbol{\Phi}$ and calculate the expectation over $\mathbf{h}_k$'s and subsequently take the expectation over $\boldsymbol{\Phi}$. The distribution of $\gamma_{k,m}$ given $\boldsymbol{\Theta}$ (stated below) is calculated using an approach similar to the one in \cite{tareq}, which studied the correlated Gaussian BC.

\begin{lemma}\label{LemmaCDF} The CDF, $F_s(x)$, and PDF, $f_s(x)$, of $\gamma_{k,m}$ defined in \eqref{SINR_RBF} for a given $\boldsymbol{\Phi}$ under the channel model in \eqref{ch_RS_RB} are given as \vspace{-.1in}
\begin{align}
\label{CDF_RBF}
&F_s(x)\hspace{-.04in}=\hspace{-.04in}1-\frac{\lambda_M}{\text{det}(\mathbf{R})} \left( \prod_{i=1}^{M-1} \frac{\lambda_i \lambda_M}{x(\lambda_i-\lambda_M)}\right) \exp\left(-\frac{x}{\frac{P_T}{\sigma^2 M} \lambda_M}\right), \\
&f_s(x)=\frac{1}{\text{det}(\mathbf{R})} \exp\left(-\frac{x}{\frac{P_T}{\sigma^2 M}  \lambda_M}\right) \prod_{i=1}^{M-1} \frac{\lambda_i \lambda_M}{x(\lambda_i-\lambda_M)} \nonumber \\ 
\label{PDF_RBF}
&\left(\frac{\mathbf{q}_M^H\mathbf{C}\mathbf{q}_M}{\frac{P_T}{M \sigma^2 }\lambda_M}\hspace{-.04in} - \hspace{-.04in}\mathbf{q}_M^H\mathbf{B}\mathbf{q}_M \hspace{-.02in}-\hspace{-.02in}\sum_{i=1}^M \frac{\lambda_M^2 \mathbf{q}_i^H\mathbf{C}\mathbf{q}_i  - \lambda_i^2 \mathbf{q}_M^H\mathbf{C}\mathbf{q}_M}{\lambda_i x (\lambda_i-\lambda_M)} \right),
\end{align}
where $\lambda_i$ is the $i^{th}$ eigenvalue of $\mathbf{A}_m$ ordered as $\lambda_1 < \dots < \lambda_M$, $\mathbf{q}_i$ is the associated eigenvector,  $\mathbf{A}_m=(1+x)\boldsymbol{\Lambda}^{1/2} \bar{\boldsymbol{\phi}}_m \bar{\boldsymbol{\phi}}_m^H \boldsymbol{\Lambda}^{1/2}-x\boldsymbol{\Lambda}$, $\boldsymbol{\Lambda}$ is the result of the eigenvalue decomposition of $\mathbf{R}$, i.e. $\mathbf{R}=\mathbf{U}^H \boldsymbol{\Lambda} \mathbf{U}$ , $\bar{\boldsymbol{\phi}}_m=\mathbf{U}\boldsymbol{\phi}_m$, $\mathbf{B}=\boldsymbol{\Lambda}^{1/2} (\bar{\boldsymbol{\phi}}_m \bar{\boldsymbol{\phi}}_m^H -\mathbf{I}_M) \boldsymbol{\Lambda}^{1/2}$ and $\mathbf{C}=\boldsymbol{\Lambda}^{1/2} \bar{\boldsymbol{\phi}}_m \bar{\boldsymbol{\phi}}_m^H  \boldsymbol{\Lambda}^{1/2}$. 
\end{lemma}

Using the derived CDF and PDF of SINR, we develop the scaling law of $\underset{k=1,\dots,K}{\text{max}}\gamma_{k,m}$ given $\boldsymbol{\Phi}$. Since $\gamma_{k,m}$ are i.i.d. across $k$ (for a given $\boldsymbol{\Phi}$) due to the channels $\mathbf{h}_k$ being i.i.d. across $k$ as discussed after Assumption 1, so the only condition remaining to be checked to apply Lemma 1 on \eqref{exp1} is whether the growth function satisfies (\ref{growth}). For this, we require the following result.
\begin{lemma}\label{LemmaLamb} The maximum eigenvalue of $\textbf{A}_m(x)$, denoted as $\lambda_{M}(x)$, and the corresponding eigenvector $\mathbf{q}_M(x)$ are given by
\begin{align}
\label{lamb}
&\lambda_M(x)=\frac{1}{\bar{\boldsymbol{\phi}}_m^H \boldsymbol{\Lambda}^{-1}\bar{\boldsymbol{\phi}}_m }, \hspace{.05in} \mathbf{q}_M(x)=\frac{\boldsymbol{\Lambda}^{-1/2} \bar{\boldsymbol{\phi}}_m}{\sqrt{\bar{\boldsymbol{\phi}}_m^H \boldsymbol{\Lambda}^{-1}\bar{\boldsymbol{\phi}}_m}}.
\end{align}
\end{lemma}
\begin{IEEEproof}
The proof is provided in Appendix \ref{App:Lamb}.
\end{IEEEproof}

Note that although $\mathbf{A}_m$ is a function of $x$, its maximum eigenvalue and corresponding eigenvector are not. Using this result, we can verify the condition in \eqref{growth} as follows.
\begin{lemma}\label{Lemmagr}
The distribution of $\gamma_{k,m}$ given $\boldsymbol{\Theta}$ satisfies the condition \eqref{growth} in Lemma \ref{Lemma1} as follows.
\begin{align}
&\underset{x\rightarrow \infty}{\text{lim}} g(x)=\underset{x\rightarrow \infty}{\text{lim}} \frac{1-F_s(x)}{f_s(x)}=\frac{P_T}{ \sigma^2 M\bar{\boldsymbol{\phi}}_m^H \boldsymbol{\Lambda}^{-1} \bar{\boldsymbol{\phi}}_m} \geq 0.
\end{align}
\end{lemma}
\begin{IEEEproof}
The proof follows from using \eqref{CDF_RBF} and \eqref{PDF_RBF} to obtain
\begin{align}
&g(x)=\frac{\lambda_M}{\frac{\mathbf{q}_M^H\mathbf{C}\mathbf{q}_M}{\frac{P_T}{M \sigma^2 } \lambda_M} - \mathbf{q}_M^H\mathbf{B}\mathbf{q}_M -\sum_{i=1}^M \frac{\lambda_M^2 \mathbf{q}_i^H\mathbf{C}\mathbf{q}_i - \lambda_i^2 \mathbf{q}_M^H\mathbf{C}\mathbf{q}_M}{\lambda_i x (\lambda_i-\lambda_M)}}.\nonumber
\end{align}
It is obvious from the definition of $\mathbf{A}$ in Lemma \ref{LemmaCDF} that $\underset{x\rightarrow \infty}{\text{lim}}\lambda_i=-\infty$ for $i\neq M$ (we showed that $\lambda_M$ is independent of $x$ in Lemma \ref{LemmaLamb}). Therefore $\underset{x\rightarrow \infty}{\text{lim}}-\frac{\lambda_M^2 \mathbf{q}_i^H\mathbf{C}\mathbf{q}_i}{\lambda_i x (\lambda_i-\lambda_M)}=0$ and $\underset{x\rightarrow \infty}{\text{lim}}-\frac{\lambda_i^2 \mathbf{q}_M^H\mathbf{C}\mathbf{q}_M}{\lambda_i x (\lambda_i-\lambda_M)}=0$. Using the definition of $\lambda_M$ and $\mathbf{q}_M$ from Lemma \ref{LemmaLamb} and that of $\mathbf{B}$ from Lemma \ref{LemmaCDF}, we can obtain $\underset{x\rightarrow \infty}{\text{lim}}\mathbf{q}_M^H\mathbf{B}\mathbf{q}_M=0$. The only non-zero term is $\frac{\lambda_M^2 \frac{P_T}{M \sigma^2 }}{\mathbf{q}_M^H\mathbf{C}\mathbf{q}_M}$ which is simplified to obtain the result.
\end{IEEEproof}

Lemma \ref{Lemmagr} therefore implies that $\underset{k=1,\dots,K}{\text{max}}\gamma_{k,m}-l_{K,m}$ converges in distribution to a limiting RV, where $l_{K,m}$ can be found using Lemma \ref{Lemma1}. The resulting scaling law is stated below.

\begin{theorem}\label{Thm2} Consider the Gaussian BC in (\ref{ch_RS_RB}) comprising of a BS with $M$ antennas, an IRS with $N$ elements and $K$ single-antenna users. Then under \textit{Assumption 1}, with RBF at the BS and random uniform phase allocation at the IRS, the sum average rate scales as
\begin{align}
&R_{RBF}^{(K)}=M \log ((\beta_r\sum_{n=1}^N \alpha_n^2 +\beta_d) \log K) + M\log \left(\frac{P_T}{\sigma^2 M}\right) \nonumber \\
&+ \sum_{m=1}^M \mathbb{E}_{\boldsymbol{\phi}_m}\left[ \log \frac{ 1}{\boldsymbol{\phi}_m^H \bar{\mathbf{R}}^{-1}\boldsymbol{\phi}_m }\right]+o(1),\nonumber
\end{align}
for large $K$, where $\bar{\mathbf{R}}=\frac{1}{\beta_r \sum_{n=1}^N \alpha_n^2  +\beta_d}\mathbf{R}$.
\end{theorem}
\begin{IEEEproof}
The proof can be found in Appendix \ref{App:Th2}.
\end{IEEEproof}

\begin{remark}\label{rem} The implementation of the scenario described in Theorem 2 is as follows. At the start of each coherence interval, the $M$-antenna BS generates $M$ random orthonormal i.d. beamforming vectors $\boldsymbol{\phi_m} \in \mathbb{C}^{M\times 1}$, $m=1,\dots, M$ according to the Gram-Schmidt decomposition method described at the start of Sec. III-B, while the  IRS applies a set of $N$ phase shifts $\theta_n$, $n=1,\dots, N$ across the $N$ reflecting elements, which are drawn from the uniform distribution over $[0,2\pi]$. The IRS continues to apply the same set of phase shifts during the training phase and the downlink transmission phase of the coherence interval. During the training phase, the BS transmits $M$ pilot symbols on the $M$ random beams (i.e. on $\boldsymbol{\phi}_m$'s) and each user measures the $M$ downlink SINRs $\gamma_{k,m}$ defined in \eqref{SINR_RBF}. Each user then feeds back its maximum SINR i.e. $\underset{m=1,\dots, M}{\text{max}} \gamma_{k,m}$, along with the beam index $m$  on which the SINR is maximized. Once the BS has this partial CSI, it assigns the beamforming vector $\boldsymbol{\phi}_m$ to transmit data symbols to the user with the highest corresponding SINR, i.e. on beamforming vector $\boldsymbol{\phi}_m$ transmit to user $k'=\underset{k=1,\dots, K}{\text{arg max}} \gamma_{k,m}$. Using this scheduling, the BS constructs the signal vector in \eqref{sig} and starts data transmission. After the coherence interval of $T$ channel uses, the BS chooses another set of beamforming vectors $\boldsymbol{\phi_m}$'s and IRS applies another set of uniformly drawn random phase shifts and the process is repeated.   The sum average rate scaling for this framework is provided in Theorem \ref{Thm2}.
\end{remark}

To complete the result in Theorem \ref{Thm2}, it remains to calculate the expectation $\mathbb{E}_{\boldsymbol{\phi}_m}\left[ \log \frac{ 1}{\boldsymbol{\phi}_m^H \bar{\mathbf{R}}^{-1}\boldsymbol{\phi}_m }\right]$.  The final scaling law after evaluating this expectation is given in Theorem \ref{Cor1} below.

\begin{theorem} \label{Cor1} Under the setting of Theorem \ref{Thm2},  the sum average rate scaling of RBF  is given as
\begin{align}
\label{Th_RBFF}
&\hspace{-.08in}R_{RBF}^{(K)}\hspace{-.04in}=\hspace{-.04in}M\log \left(\frac{P_T}{\sigma^2 M}\right) \hspace{-.04in}+\hspace{-.04in}M \log ((\beta_r\sum_{n=1}^N \alpha_n^2 +\beta_d) \log K) \nonumber \\
&+ M \log(\lambda_1(\bar{\boldsymbol{\Lambda}}))\hspace{-.04in} + \hspace{-.04in}M\sum_{i=1}^N \eta_i  \left(\hspace{-.04in}\frac{-1}{\lambda_i(\bar{\boldsymbol{\Lambda}})} \hspace{-.04in} \right)^{M-1}\log \left( \frac{\lambda_i(\bar{\boldsymbol{\Lambda}})}{\lambda_1(\bar{\boldsymbol{\Lambda}})} \right) \nonumber \\
&+M \sum_{i=1}^N \eta_i \sum_{l=1}^{M-1} \binom{M-1}{l}\frac{1}{l}\left(\frac{1}{\lambda_1(\bar{\boldsymbol{\Lambda}})^l}-\frac{1}{\lambda_i(\bar{\boldsymbol{\Lambda}})^l}\right)\nonumber \\
&\left( -\frac{1}{\lambda_i(\bar{\boldsymbol{\Lambda}})} \right)^{M-1-l} \hspace{-.1in} +o(1),
\end{align} 
\end{theorem}
where $\bar{\mathbf{R}}=\mathbf{U}^H \bar{\boldsymbol{\Lambda}} \mathbf{U}$,  $\lambda_i(\bar{\boldsymbol{\Lambda}})$ is the $i^{th}$ diagonal element of $\bar{\boldsymbol{\Lambda}}$ and $\lambda_1(\bar{\boldsymbol{\Lambda}}) <\dots < \lambda_M(\bar{\boldsymbol{\Lambda}})$.
\begin{IEEEproof}
The expression of $\mathbb{E}_{\boldsymbol{\phi}_m}\left[ \log \frac{ 1}{\boldsymbol{\phi}_m^H \bar{\mathbf{R}}^{-1}\boldsymbol{\phi}_m }\right]$ is derived in Appendix \ref{App:Lemma2}. Plugging this expression in Theorem \ref{Thm2} yields Theorem \ref{Cor1}.
\end{IEEEproof}

The expression is involved but yields some insights. There is an increase in the sum average rate as compared to \eqref{RBF_sc} by a factor of $M\log \left(\frac{\beta_r}{\beta_d}\sum_{n=1}^N \alpha_n^2 +1\right)$. However, there is also a decrease corresponding to $\sum_{m=1}^M \mathbb{E}_{\boldsymbol{\phi}_m}\left[ \log \frac{ 1}{\boldsymbol{\phi}_m^H \bar{\mathbf{R}}^{-1}\boldsymbol{\phi}_m }\right]$, due to the correlated nature of the channel.

\begin{remark}
The analysis can be extended to the scenario with $L$ antennas at each user. A direct extension is to treat each receive antenna as an independent user and have effectively $LK$ single antenna users \cite{RBF}.  Therefore, each user will feed back $L$ times the amount of information because corresponding to each receive antenna $i$, $i=1,\dots, L$ the user needs to feed back the maximum SINR, i.e. $\underset{m=1,\dots, M}{\text{arg max}} \gamma_{i,m}$ along with the maximizing beam index $m$. The BS then assigns beamforming vector $\boldsymbol{\phi}_m$, $m=1,\dots, M$ to transmit data symbols to the antenna of the user with the highest SINR, i.e., $\underset{i=1,\dots, LK}{\text{arg max}} \gamma_{i,m}$. Since the maximization is over $LK$ i.i.d. RVs, the scaling law will be the same as the one in Theorem \ref{Cor1} with $\log K$  replaced by $\log (LK)$.

The results can be extended, using some involved analysis, to the case where at most one beam is assigned to each user by computing the overall SINR of each user instead of the SINR for each receive antenna separately. However, this is beyond the scope of this work.
\end{remark}

\subsection{Deterministic Beamforming}
\label{Sec:Asym3}
We also study the case where the matrix $\boldsymbol{\Phi}$ is fixed over all channel uses and refer to this scheme as DBF \cite{tareq}. The analysis stays the same as that for RBF, except that we do not need to take the expectation over $\boldsymbol{\Phi}$ in Theorem \ref{Thm2}. The sum average rate is then written as \vspace{-.05in}
\begin{align}
&R_{DBF}^{(K)}=M\log \left(\frac{P_T}{\sigma^2 M}\right) +M \log ((\beta_r\sum_{n=1}^N\alpha_n^2+\beta_d) \log K) \nonumber \\
&+  \sum_{m=1}^M \log \frac{ 1}{\bar{\boldsymbol{\phi}}_m^H \bar{\boldsymbol{\Lambda}}^{-1}\bar{\boldsymbol{\phi}}_m }+o(1),
\end{align}
for large $K$, where $\bar{\boldsymbol{\phi}}_m=\mathbf{U} \boldsymbol{\phi}_m$ and $\mathbf{U}^H \bar{\boldsymbol{\Lambda}}^{-1} \mathbf{U}$ is the eigenvalue decomposition of $\bar{\mathbf{R}}^{-1}$. An interesting special case will be when $\mathbf{U} \boldsymbol{\phi}_m$ are columns of the identity matrix. In this case, the DBF matrix $\boldsymbol{\Phi}$ is equal to $\mathbf{U}^H$ and therefore $\frac{ 1}{\bar{\boldsymbol{\phi}}_m^H \bar{\boldsymbol{\Lambda}}^{-1}\bar{\boldsymbol{\phi}}_m }=\lambda_m(\bar{\mathbf{R}})$. This leads to $\sum_{m=1}^M \log \frac{ 1}{\bar{\boldsymbol{\phi}}_m^H \bar{\boldsymbol{\Lambda}}^{-1}\bar{\boldsymbol{\phi}}_m }=\sum_{m=1}^M \log \lambda_m(\bar{\mathbf{R}})=\log \text{det} (\bar{\mathbf{R}})$. Thus we obtain the following result.

\begin{theorem}\label{Corr2}  Consider the Gaussian BC in (\ref{ch_RS_RB}) comprising of a BS with $M$ antennas, an IRS with $N$ elements and $K$ single-antenna users. Then under \textit{Assumption 1}, the DBF scheme at the BS and random phase rotations at the IRS, the sum average rate scales  as \vspace{-.1in}
\begin{align}
\label{Cor2}
&R_{DBF}^{(K)}=M\log \left(\frac{P_T}{\sigma^2 M}\right) + M \log ((\beta_r\sum_{n=1}^N \alpha_n^2+\beta_d) \log K )\nonumber \\
&+ M \log (\text{det}(\bar{\mathbf{R}}))^{1/M}+o(1).
\end{align} 
\end{theorem}
As compared to the sum average rate scaling in the conventional MISO BC given by \eqref{RBF_sc}, we see that the random-rotations IRS scheme increases the sum-rate by $M\log \left(\frac{\beta_r}{\beta_d} \sum_{n=1}^N \alpha_n^2+1\right)$ and decreases it by $|\log (\text{det } \bar{\mathbf{R}})|$ since $\text{det } \bar{\mathbf{R}}\leq 1$. This also proves that the right hand side of the upper bound on the average sum-capacity scaling in \eqref{temp33} is achievable, and thus \eqref{Cor2} acts as a lower bound to the average sum-capacity scaling and completes the proof of Theorem \ref{Thm1}.

\subsection{Complexity and CSI Overhead Comparison}

So far we have developed the scaling laws of the average sum-capacity and the sum average rate for different transmission schemes at the BS while employing the random rotations scheme at the IRS. The main motivation behind using the random rotations scheme at the IRS is as follows. Unlike the coherent beamforming scheme where the IRS phase shifts need to be optimized based on the instantaneous CSI of IRS-user ($\mathbf{h}_{2,k}\in \mathbb{C}^{N\times 1}$) and BS-user ($\mathbf{h}_{d,k} \in \mathbb{C}^{M\times 1}$) channels, the proposed scheme does not require the availability of the instantaneous CSI of these individual IRS-assisted and direct channels to implement the random phase shifts. The only CSI needed is, therefore, at the BS of the overall channel $\mathbf{h}_k \in \mathbb{C}^{M\times 1}$ to implement the beamforming transmission scheme. This saves  the large training overhead associated with the estimation of $\mathbf{h}_{2,k}$'s and $\mathbf{h}_{d,k}$'s \cite{annie_OJ}. Also, the random rotations scheme reduces the system complexity since optimizing IRS-phase shifts at the pace of fast-fading channels is not required. 

In this part, we discuss the computational complexity and CSI feedback overhead associated with the beamforming schemes considered at the BS. The sum-capacity of a Gaussian BC is achieved by  DPC \cite{costa}, based on the idea that when the  interference caused by other users' signals is known at the BS in advance (non-causally), it is possible to achieve the same capacity as if there was no interference by successively encoding the users while preserving the power constraint. DPC is well-known to be extremely computationally intensive to implement due to the high computational burden of successive encoding and decoding \cite{ZF_GUS}. The exact complexity in terms of the number of complex arithmetic operations largely depends on the implementation of the DPC based optimal algorithm.  An efficient iterative algorithm was proposed by  Jindal \textit{et al} in \cite{jindal}, which requires $O(M^2 K^2)$ operations \textit{per iteration}. To circumvent this problem, many works advocate the use of the linear zero-forcing (ZF) beamforming  to create orthogonal channels between the BS and users. Since in our setting, $K>>M$, so we consider ZF with user selection (ZFS) from \cite{ZF_GUS} as a benchmark scheme in the simulations, wherein the user selection procedure has a computational complexity of $O(M^3 K)$ while ZF with $M$ selected users has a complexity of $O(M^3)$  \cite{ZFabla}. Both DPC based algorithms and ZF require full CSI feedback from all users to the BS. Therefore, each user needs to feedback $2M$ real numbers corresponding to $\mathbf{h}_{k}\in \mathbb{C}^{M\times 1}$ to the BS at the start of each coherence interval.

We then studied two low-complexity beamforming schemes that require partial CSI at the BS. Both RBF and DBF require each user to feedback only one real number (its maximum SINR) and the corresponding beam index (an integer) to the BS at the start of each coherence interval, which is significantly less than the CSI overhead associated with DPC-based and ZF schemes.  In terms of implementation, RBF requires $M$ random orthonormal  beams $\boldsymbol{\phi}_m$'s to be generated according to an isotropic distribution, which has a computational complexity of $O(M^3)$ per coherence interval. DBF requires the beamforming matrix $\boldsymbol{\Phi}$ to be set as $\mathbf{U}^H$ obtained from the eigenvalue decomposition of $\bar{\mathbf{R}}^{-1}$, which also has a computational complexity of $O(M^3)$. However this computation does not need to be performed per coherence interval but only when the large-scale channel statistics change after several coherence intervals. The user selection algorithm (i.e. finding the user with maximum SINR on each beam) at the BS for both RBF and DBF has a computational complexity of $O(K M)$, which is significantly less than that of ZFS.

Overall, both RBF and DBF impose a significantly lower CSI feedback overhead than DPC and ZFS. The associated user scheduling scheme for RBF and DBF also has a reduced complexity. These observations are summarized in Table \ref{tablecom}, along with the computational complexity and CSI feedback results for the random rotations IRS scheme and the benchmark coherent IRS beamforming scheme (implemented in simulations using exhaustive search with $Q$ phase shift levels).

\begin{table*}[!b]
\centering
\normalsize
\caption{Computational complexity and feedback overhead comparison of considered beamforming schemes at the BS and IRS. All values are provided per coherence interval.}
\begin{tabular}{|l|l|l|l|}
\hline
\centering
  \textbf{Scheme} & \textbf{Feedback  }& \textbf{Complexity of Beamforming} & \textbf{Complexity of User} \\ 
	& (in real numbers) & (for $M$ selected users) & \textbf{ Selection} \\
\hline
	\textbf{Base Station} & & & \\
ZFS \cite{ZF_GUS} & $2KM$ & $O(M^3)$ & $O(M^3 K)$ \\
RBF & $2K$ \cite{RBF} & $O(M^3)$ & $O(MK)$ \\
DBF & $2K$ \cite{RBF} & $O(M^3)$  & $O(MK)$ \\
\hline
	\textbf{IRS} & & & \\
Random rotations & \text{None} & $O(N)$  &  -\\
Coherent BF & $2KN+2KM $& $O(Q^{N})$ & -  \\
\hline
\end{tabular}
\label{tablecom}
\end{table*}

\section{Energy Efficiency Optimization}
\label{Sec:Opt}

In this section, we optimize the IRS-assisted MISO BC in terms of EE by using the sum-rate results developed in last section. Current works that study the EE of IRS-assisted MISO BC consider coherent beamforming schemes at the IRS under perfect CSI assumption.  Different from these works, we consider the random rotations scheme at the IRS and DBF (or RBF) at the BS. Under this framework, no CSI is needed at the IRS since it just introduces random phase-shifts from the uniform distribution and  only partial CSI  is needed at the BS to implement DBF or RBF. Under this setting, we aim to obtain the optimal system configuration parameters, including the numbers of BS antennas $M$ and IRS elements $N$ and the Tx power $P_T$, that maximize EE. The main analysis is presented for DBF at the BS and extends straightforwardly to RBF as well.

\subsection{Problem Formulation}

We first describe the power consumption model, which will consist of the Tx power $P_T$ at the BS, the circuit power $P_{c}$ consumed at the BS and users, the static power $P_{s,B}$ consumed at the BS and the power  $P_{IRS}$ consumed at the IRS.  The total power consumed is given as
\begin{align}
&P_{tot}=(1/\zeta) P_T + P_{c}+P_{s,B}+P_{IRS},
\end{align} 
where $\zeta$ is the power amplifier efficiency. The circuit power model is given as \cite{power}
\begin{align}
&P_{c}=\underbrace{M(P_{DAC}+P_{mix}+P_{filt})}_{\text{Consumed by the BS}}\nonumber \\
&+\underbrace{M(P_{LNA}+P_{mix}+P_{filr}+P_{ADC})}_{\text{Consumed by the $M$ users}},
\end{align} 
where $P_{mix}$, $P_{filt}$, $P_{filr}$, $P_{LNA}$, $P_{DAC}$ and $P_{ADC}$ represent the power consumed in the mixer, filters of transmitter and receiver, low noise
amplifier, digital-to-analog converter (DAC) and analog-to-digital converter (ADC), respectively. The IRS power consumption is given as $P_{IRS}=N P_{n}+P_{s,I}$, where $P_{n}$ is the power consumed by one IRS element and $P_{s,I}$ is the static power consumption of the IRS. The total power can then be written in  a compact form as
\begin{align}
\label{tot}
&P_{tot}=AP_{T}+BM + CN+ D,
\end{align}
where $A=\frac{1}{\zeta}$, $B=P_B+P_U$, $P_B=P_{DAC}+P_{mix}+P_{filt}$, $P_U=P_{LNA}+P_{mix}+P_{filr}+P_{ADC}$, $C=P_n$ and $D=P_{s,B}+P_{s,I}$.

Since the focus of this work is on the large $K$ regime, we utilize the sum average rate scaling expression $R_{DBF}^{(K)}$ derived in \eqref{Cor2} to formulate the EE scaling expression under DBF as
\begin{align}
\label{EE}
&\text{EE}^{(K)}=\frac{\splitdfrac{M \log ((\beta_r\sum_{n=1}^N \alpha_n^2 +\beta_d) \log K) + M\log\left( \frac{P_T}{\sigma^2 M}\right)}{ + \log \det(\bar{\mathbf{R}})}}{AP_{T}+BM + CN+ D},
\end{align}
where $\bar{\mathbf{R}}=\frac{1}{ \beta_r\sum_{n=1}^N \alpha_n^2+\beta_d} (\beta_r \mathbf{H}_1 \text{diag}(\alpha_1^2, \dots, \alpha_N^2) \mathbf{H}_1^H + \beta_d \mathbf{I}_M)$. 

To determine the optimal values of the number of BS antennas $M^*$, the number of IRS elements $N^*$, and the optimal power $P_T^*$, we consider the following problem:
 \begin{align} \textit{(P1)} \hspace{.25in} \max_{M, N, P_T}      &\frac{\splitdfrac{M \log ((\beta_r\sum_{n=1}^N \alpha_n^2 +\beta_d) \log K)} {+ M\log\left( \frac{P_T}{\sigma^2 M}\right) + \log \det(\bar{\mathbf{R}})}}{AP_{T}+BM + CN+ D}, \label{P1}
\end{align}
subject to constraints on the maximum number of deployable transmit antennas and IRS reflecting elements, i.e. $M\in{\cal M}=\{1,\ldots,M_{max}\}$ and $N\in{\cal N}=\{1,\ldots,N_{max}\}$, and on the maximum feasible Tx power, i.e. $P_{T}\leq P_{max}$.

To explicitly show the dependence of $\det(\bar{\mathbf{R}})$ on $M$ and $N$, we express the determinant of $\bar{\mathbf{R}}\in \mathbb{C}^{M\times M}$ using Leibniz formula and the definition of $\mathbf{H}_1$ in \eqref{H_1} to obtain $\det(\bar{\mathbf{R}})=$
\begin{align}
 &\hspace{-.03in}\sum_{\underset{ \boldsymbol{\sigma}_i \in \mathbf{S}_M}{i=1}}^{M!} \hspace{-.03in} sgn(\boldsymbol{\sigma}_i)\prod_{m=1}^M \hspace{-.03in} \frac{ \splitdfrac{\beta_r \sum_{n=1}^N \alpha_n^2 \exp\Big(j \frac{2\pi}{\lambda} (l_{(m), (n_1,n_2)}}{-\l_{(\boldsymbol{\sigma}_i(m)),(n_1,n_2)})\Big)+\beta_d\mathbf{I}_{M}(m,\boldsymbol{\sigma}_i(m))}}{\beta_r\sum_{n=1}^N\alpha_n^2 +\beta_d }, \label{P2}
\end{align}
where $\mathbf{S}_M\in \mathbb{R}^{M\times M!}$ represents the set of all $M!$ permutations of $[1, \dots, M]$, where each permutation denoted as $\boldsymbol{\sigma}_i \in \mathbb{R}^{M\times 1}$ is the $i^{th}$ column of $\mathbf{S}_M$. Moreover $sgn(\boldsymbol{\sigma}_i)$ denotes the signature of $\boldsymbol{\sigma}_i$ which will be $1$ if $\boldsymbol{\sigma}_i$ is an even permutation and $-1$ if $\boldsymbol{\sigma}_i$ is an odd permutation.

 Finding the optimal solution of \eqref{P1} with affordable complexity is complicated due to the fact that two of the three variables are discrete and also that both $M$ and $N$ appear as the upper limits of sums and products terms as seen in \eqref{P2}, which prevents us from using standard gradient-based methods. For these reasons, we will consider both the maximization of the exact objective of \eqref{P1} by line searches methods, and the maximization of a bound of this objective. 

Note that the problem \textit{(P1)} can be similarly formulated under RBF by using the sum-average rate scaling $R_{RBF}^{(K)}$, derived in \eqref{Th_RBFF}, as the numerator of $\text{EE}^{(K)}$ in \eqref{P1}.

\subsection{Exact Solution of Problem \textit{(P1)}}
The global solution of Problem \textit{(P1)} can be determined by an exhaustive search in the set ${\cal S}=\{{\cal M}\times {\cal N}\times [0,P_{max}]\}$. Using a step size $\delta$ to search the continuous set $[0,P_{max}]$, this approach requires $M_{max}N_{max}P_{max}/\delta$ computations and comparisons of the values of the objective function of \eqref{P1}. Although globally optimal, this technique might be too demanding for large values of $M_{max}$ and $N_{max}$, or for low values of $\delta$. 

A first approach to reduce the computational complexity is based on the use of alternating maximization to optimize \eqref{P1} with respect to $M$, $N$, and $P_{T}$ one variable at a time, while keeping the other two fixed. Formally, denoting by $F(M,N,P_{T})$ the objective of Problem \textit{(P1)}, an alternating maximization algorithm for Problem \textit{(P1)} can be stated in Algorithm \ref{Alg:ExactAM}.
\begin{algorithm}[!b]
\caption{Alternating Maximization for Problem \textit{(P1)}}
\label{Alg:ExactAM}
\begin{algorithmic}
\State \texttt{Set} $M=\bar{M}\in{\cal M}$, $N=\bar{N}\in{\cal N}$, $P_{T}=\bar{P}_{T}\in[0,P_{max}]$; $F=F(\bar{M},\bar{N},\bar{P}_{T})$;
\While{$|\bar{F}-F|\geq \varepsilon$}
\State $\bar{F}=F$;
\State $\bar{P}_{T}=\text{arg max}_{P_{T}}\;F(\bar{M},\bar{N},P_{T})$;
\State $\bar{M}=\text{arg max}_{M}\;F(M,\bar{N},\bar{P}_{T})$;
\State $\bar{N}=\text{arg max}_{N}\;F(\bar{M},N,\bar{P}_{T})$;
\State $F=F(\bar{M},\bar{N},\bar{P}_{T})$;
\EndWhile
\State \texttt{Output} $\bar{M}$, $\bar{N}$, $\bar{P}_T$
\end{algorithmic}
\end{algorithm}

The convergence of Algorithm \ref{Alg:ExactAM} is ensured by the fact that at each step the objective function does not decrease. Since EE can not grow to infinity, Algorithm \ref{Alg:ExactAM} must eventually converge in the value of the objective. As for its complexity, Algorithm \ref{Alg:ExactAM} requires two exhaustive searches in each iteration to optimize $M$ and $N$, while the optimization with respect to $P_{T}$ can be carried out in a semi-closed-form, as shown next.
\begin{theorem}\label{Th:EEmaxPT}
For any fixed $M$ and $N$, the objective function of Problem (\textit{P1})  admits a unique maximizer with respect to $P_{T}$, which is given by 
\begin{equation}\label{Eq:OptPt}
\bar{P}_{T}=\min\{P_{T}^{\star},P_{max}\},
\end{equation}
where $P_T^{\star}$ is obtained as the root of 
\begin{equation}\label{Eq:StPoint}
d+\frac{c}{P_{T}}- db-d\log (aP_{T})=0.
\end{equation}
\end{theorem}
\begin{IEEEproof}
Define $a=\frac{1}{\sigma^2 M}$, $b=\log ((\beta_r\sum_{n=1}^N\alpha_n^2 +\beta_d) \log K)+\frac{1}{M}\log \det(\bar{\mathbf{R}})$, $c=\frac{BM + CN+ D}{M}$ and $d=\frac{A}{M}$. The objective function in \eqref{P1} can be written as a function of $P_{T}$ as 
\begin{equation}\label{Eq:EE}
\text{EE}^{(K)}=\frac{\log(aP_{T})+b}{dP_{T}+c}\;.
\end{equation}
Then, the first-order derivative of \eqref{Eq:EE} is positive whenever 
\begin{equation}
\label{con_der}
d+\frac{c}{P_{T}}\geq db+d\log (aP_{T})\;,
\end{equation}
which is verified for $P_{T}\to 0$ and not verified for $P_{T}\to\infty$, implying that the function \eqref{Eq:EE} has at least one stationary point.

Next, let us observe that the objective \eqref{Eq:EE} as a function of only $P_{T}$, has a strictly concave numerator and an affine denominator. As a result, it is a strictly pseudo-concave function, being the ratio between a strictly concave function and an affine function. Strictly pseudo-concave functions are known to be either monotonically increasing or to admit a unique stationary point, which coincides with the function's global maximizer. As for the objective in \eqref{Eq:EE}, there exists a unique stationary point, say $P_{T}^{\star}$, because we have found the condition in \eqref{con_der} to be verified for $P_{T}\to 0$ and not for $P_{T}\to\infty$. The theorem then follows from the constraint $P_{T}\leq P_{max}$.
\end{IEEEproof}

Denoting by $I$ the number of iterations until convergence, Algorithm \ref{Alg:ExactAM} requires $IM_{max}N_{max}$ evaluations and comparisons of the objective of Problem \textit{(P1)}, plus $I$ evaluations of \eqref{Eq:OptPt}. While the number of iterations $I$ for convergence is not known in advance, it is typically of the order of a few units. Moreover, $P_{T}^{\star}$ in \eqref{Eq:OptPt} can be conveniently evaluated using standard numerical methods for solving transcendental equations, e.g. bisection search applied to \eqref{Eq:StPoint}.

\begin{remark}
The global solution under RBF can be similarly determined by an exhaustive search in the set ${\cal S}=\{{\cal M}\times {\cal N}\times [0,P_{max}]\}$, where the objective function evaluated in \textit{(P1)} is $\frac{R_{RBF}^{(K)}}{A P_T +BM+CN+D}$, where $R_{RBF}^{(K)}$ is given in \eqref{Th_RBFF}. Similarly the alternating optimization in Algorithm 1 will be the same for RBF as well, with the objective of \textit{(P1)}, denoted by $F(M,N,P_T)$, given as $\frac{R_{RBF}^{(K)}}{A P_T +BM+CN+D}$. The definition of $b$ in Theorem \ref{Th:EEmaxPT} to find $P_T^*$ will be $b=\log ((\beta_r\sum_{n=1}^N\alpha_n^2 +\beta_d) \log K)+\mathbb{E}_{\boldsymbol{\phi}}\left[ \log \frac{ 1}{\boldsymbol{\phi}_m^H \bar{\mathbf{R}}^{-1}\boldsymbol{\phi}_m }\right]$, where the expression of $\mathbb{E}_{\boldsymbol{\phi}}\left[ \log \frac{ 1}{\boldsymbol{\phi}_m^H \bar{\mathbf{R}}^{-1}\boldsymbol{\phi}_m }\right]$ has been derived in Appendix \ref{App:Lemma2}.
\end{remark}

\subsection{Low-Complexity Solution of Problem \textit{(P1)}}
The methods just developed  involve an exhaustive search either in two or three dimensions. Instead, in this section we optimize an upper bound on the objective function of Problem \textit{(P1)} that does not require any exhaustive search. To this end, we proceed as follows.

Note that in the objective function in \eqref{P1}, $\log(\det(\bar{\mathbf{R}}) )\leq 0$ with equality when $\bar{\mathbf{R}}=\mathbf{I}_M$. This is because the value of $\det(\bar{\mathbf{R}})$ is between $0$ and $1$,  as discussed in Sec. III. The value of $\log \det(\bar{\mathbf{R}})$ will therefore determine the rate loss caused by the correlation introduced into the channel by the BS-IRS LoS link $\mathbf{H}_1$.  This observation allows us to upper bound the EE scaling under DBF (as well as RBF) under the assumption that $\alpha_n=\alpha$, $\forall n$\footnote{Note that the assumption $\alpha_n=\alpha$, $\forall n$ is made in almost all existing papers on IRS-assisted systems since the reflecting elements in a single IRS will generally have the same construction.} as 
\begin{align}
\label{bounbd}
&\text{EE}^{(K)}\leq \frac{M \log ((N\alpha^2 \beta_r+\beta_d) \log K) + M\log \left(\frac{P_T}{\sigma^2 M}\right)}{AP_{T}+BM + CN+ D}.
\end{align}

In this part we will focus on optimizing this bound on EE instead of the exact objective function in Problem \textit{(P1)}, resulting in the following optimization problem.
 \begin{align} \textit{(P2)} \hspace{.2in}
\max_{M, N, P_T}      &  \frac{M \log ((N\alpha^2 \beta_r+\beta_d) \log K) + M\log \left(\frac{P_T}{\sigma^2 M}\right)}{AP_{T}+BM + CN+ D}, \label{P3}
\end{align}
subject to $M\in{\cal M}=\{1,\ldots,M_{max}\}$, $N\in{\cal N}=\{1,\ldots,N_{max}\}$ and $P_{T}\leq P_{max}$. 

We would stress  that the reason we are considering the upper-bound on the EE as an objective function here is to develop a low-complexity method to obtain the solutions for $M$, $N$ and $P_T$, that achieve EE values close to those yielded by the exact algorithms developed in the last subsection.  We will numerically study the true EE performance (i.e. the value of the function in \eqref{P1}) under the values of $M$, $N$ and $P_T$ obtained by solving \textit{(P2)}, in a realistic setting. Interestingly, we will see that the true EE performance under the solution of    \textit{(P2)} is close to the EE performance under the solutions of the exact Problem \textit{(P1)}. 

We would also remark here that theoretically  the upper bound on EE in \eqref{bounbd} starts to approach the exact EE in \eqref{P1} when $\mathbf{H}_1\mathbf{H}_1^H=N \mathbf{I}_M$ which results in $\bar{\mathbf{R}}=\mathbf{I}_M$ and $\log \det (\bar{\mathbf{R}})=0$. This implies that as the rows of the LoS channel matrix $\mathbf{H}_1$ in \eqref{H_1} become orthogonal, the bound starts to become tight. The criteria to achieve full orthogonality in terms of the array parameters at the BS and IRS has been derived for a URA at the IRS in \cite{losura}, and requires the distance between the BS and the IRS, denoted as $\bar{D}$, to become extremely small or the number of elements in the IRS to become very large, especially at the current mmWave frequencies. 

 Although $\mathbf{H}_1$ in \eqref{H_1} is  observed to have a high rank for small to moderate BS-IRS distances (under the considered spherical wave model), high rank does not guarantee full orthogonality. In fact, in most practical settings $\mathbf{H}_1$ will not have orthogonal rows since the condition for orthogonality discussed above will not hold, so the bound will not be tight. However, we stress that the fact that the bound is not tight does not directly impact the accuracy of the optimization. The important thing is to see if the resulting EE performance under the obtained maximizers is close irrespective of whether the functions in \eqref{P1} and \eqref{P3} are close or not. For example: the authors in \cite{bounddes} used upper and lower bounds on the rate and EE that are not tight to find the optimal IRS beamforming and other system parameters. The maximizers they obtained, when plugged into the true objective functions, yielded very good results. Since we have already provided two exact methods to solve \textit{(P1)} in the last section, here we are trying to develop a more computationally-friendly method that can be employed when computational cost is an issue. The performance gap between \textit{(P1)} and \textit{(P2)} will  reduce as $\bar{D}$ becomes smaller, since the bound becomes closer (although not fully tight) to the true EE function.

A low-complexity solution for \textit{(P2)} is now developed using alternating optimization of the variables $M$, $N$, and $P_{T}$ and noting that \eqref{P3} lends itself to a semi closed-form maximization with respect to all three variables. Let us analyze the optimization of  $M$, $N$, and $P_{T}$ separately. 
\subsubsection{Optimization of $P_{T}$, for fixed $M$ and $N$}
It can be seen that when $M$ and $N$ are fixed, \eqref{P3} can be expressed as in \eqref{Eq:EE}, with 
$a=\frac{1}{M \sigma^2}$, $b=\log ((N\alpha^2\beta_r+\beta_d) \log K)$, $c=\frac{BM + CN+ D}{M}$ and $d=\frac{A}{M}$. Then, the maximization of \eqref{P3} with respect to $P_{T}$ can be carried out following the same reasoning as that in  Theorem \ref{Th:EEmaxPT}, which leads to the maximizer 
\begin{equation}\label{Eq:OptPT}
\bar{P}_{T}=\min\{P_{max},P_{T}^{\star}\}\;,
\end{equation}
with $P_{T}^{\star}$ the unique solution of \eqref{Eq:StPoint}. 
\subsubsection{Optimization of $M$ for fixed $N$ and $P_T$}
Defining $\chi=\log ((N\alpha^2 \beta_r+\beta_d)\log K)$, $\beta=\frac{P_{T}}{\sigma^2}$, $\omega=AP_{T}+CN+D$, $\delta=B$, the objective to maximize in \eqref{P3} can be written as 
\begin{equation}\label{Eq:EE_M} 
\text{EE}^{(K)}=\frac{\chi M+M\log\frac{\beta}{M}}{\delta M+\omega}.
\end{equation}
Then, the optimal integer $M$ that maximizes \eqref{Eq:EE_M} is determined in the following theorem.
\begin{theorem}\label{Th:MaxEE_M}
The maximizer of the EE function in \eqref{Eq:EE_M} in set $\{0,1,\ldots,M_{max}\}$ is given by
\begin{equation}\label{Eq:OptM}
\bar{M}=\max\{\text{EE}(\lceil\tilde{M}\rceil),\text{EE}(\lfloor \tilde{M}\rfloor)\}\;,
\end{equation}
wherein $\tilde{M}=\min\{M_{max},M^{\star}\}$ and $M^{\star}$ is the unique solution of the equation
\begin{equation}\label{Eq:StCond}
\chi+\log\beta-1-\log(M)=\frac{\delta M(\chi +\log\beta-\log M)}{\delta M+\omega}\;,
\end{equation}
in the set $M\in[0,M_{max}]$.
\end{theorem}
\begin{IEEEproof}
In order to show the result, it is convenient to study first the maximization of \eqref{Eq:EE_M} in the continuous set $[0,M_{max}]$. We can see that the numerator of \eqref{Eq:EE_M} is a strictly concave function whereas the denominator is clearly affine with $M$. Therefore \eqref{Eq:EE_M} is a strictly pseudo-concave function. Moreover, the first-order derivative of \eqref{Eq:EE_M} is positive if
\begin{equation}\label{Eq:StCondM}
\chi+\log\beta-1-\log(M)\geq\frac{\delta M(\chi +\log\beta-\log M)}{\delta M+\omega}\;,
\end{equation}
which is verified for $M\to0$, and not verified when $M\to\infty$. Thus, \eqref{Eq:EE_M} must admit at least one stationary point, which is unique given the pseudo-concavity of \eqref{Eq:EE_M}. Then, denoting by $M^{\star}$ the unique solution of  \eqref{Eq:StCondM}, the $M$ that maximizes \eqref{Eq:EE_M} is $\tilde{M}=\min\{M_{max},M^{\star}\}$. The optimal integer $\bar{M}$ in the set $\{0,1,\ldots,M_{max}\}$ can then be found as \eqref{Eq:OptM}.
\end{IEEEproof}

\subsubsection{Optimization of $N$ for fixed $M$ and $P_{T}$}
Defining $\tau=M\log\frac{P_{T}}{\sigma^2 M}+M \log \log K$, $\gamma=AP_{T}+BM+D$, $\delta=C$, the objective to maximize in \eqref{P3} with respect to $N$ can be written as 
\begin{equation}\label{Eq:EE_N}
\text{EE}^{(K)}=\frac{M \log (N\alpha^2 \beta_r+\beta_d)+\tau}{\delta N+\gamma}.
\end{equation}
The optimal integer $N$ that maximizes \eqref{Eq:EE_N} is determined in the following theorem, whose proof follows along the same lines as that for Theorem \ref{Th:MaxEE_M}.
\begin{theorem}
The maximizer of the EE function in \eqref{Eq:EE_N} in the set $\{0,1,\ldots,N_{max}\}$ is given by
\begin{equation}\label{Eq:OptN}
\bar{N}=\max\{\text{EE}(\lceil\tilde{N}\rceil),\text{EE}(\lfloor \tilde{N}\rfloor)\}\;,
\end{equation}
wherein $\tilde{N}=\min\{N_{max},N^{\star}\}$ and $N^{\star}$ is the unique solution of the equation
\begin{equation}\label{Eq:StCond}
(\delta N +\gamma )\left(\frac{M\alpha^2 \beta_r}{N\alpha^2\beta_r+\beta_d} \right)=\delta M \log (N\alpha^2\beta_r +\beta_d)+\delta \tau.
\end{equation}
\end{theorem}

Thus, denoting by $F_{u}$ the objective function of Problem \textit{(P2)}, an alternating maximization algorithm  is formally stated in Algorithm \ref{Alg:ApproxAM}, which does not require any exhaustive search. The performance of this low-complexity algorithm as compared to the exact methods will be numerically studied in the next section by evaluating the true EE in \eqref{P1} using the values of $\bar{M}$, $\bar{N}$ and $\bar{P}_T$ obtained using Algorithm 2.

\begin{algorithm}
\caption{Alternating Maximization for Problem \textit{(P2)}}
\label{Alg:ApproxAM}
\begin{algorithmic}
\State \texttt{Set} $M=\bar{M}\in{\cal M}$, $N=\bar{N}\in{\cal N}$, $P_{T}=\bar{P}_{T}\in[0,P_{max}]$; $F_{u}=F_{u}(\bar{M},\bar{N},\bar{P}_{T})$;
\While{$|\bar{F}_{u}-F_{u}|\geq \varepsilon$}
\State $\bar{F}_{u}=F_{u}$;
\State \texttt{Set} $\bar{P}$ \texttt{as in} \eqref{Eq:OptPT};  \texttt{Set} $\bar{M}$ \texttt{as in} \eqref{Eq:OptM}; \texttt{Set} $\bar{N}$ \texttt{as in} \eqref{Eq:OptN};
\State $F_{u}=F_{u}(\bar{M},\bar{N},\bar{P}_{T})$;
\EndWhile
\end{algorithmic}
\end{algorithm}

\begin{table*}[!b]
\centering
\normalsize
\caption{Simulation parameters.}
\begin{tabular}{|l|l|l|l|}
\hline
  \textbf{Parameter} & \textbf{Value}& \textbf{Parameter} & \textbf{Value} \\ 
\hline
BS position & $(0,0)$ &IRS position & $(0, 50\rm{m})$ \\
Users positions & $(x,y)\in[-30,30]\times[50,130]$ &Bandwidth & $20$\rm{MHz} \\
Noise level $\sigma^2$ & $-80$\rm{dBm}  & $N_1$ & $8$ \\
Path loss model & $\frac{10^{-C/10}}{d^{\alpha}}$ & $C$  & $30$\rm{dB}\\
$\alpha$ & $2.2$ ($\beta_{1}$), $2.8$ ($\beta_{2}$), $3.5$ ($\beta_{d}$) \cite{LIS} & $\mathbf{H}_1$ & \eqref{H_1} \\
$\theta_{BS}, \theta_{IRS}$ & $0$ & $\phi_{BS}, \phi_{IRS}$ & $\frac{\pi}{2}$ \\
$d_{BS}$, $d_{IRS}^{(1)}$, $d_{IRS}^{(2)}$& $\lambda$ & $P_{max}$& $10$\rm{dBW} \\
$M_{max}$, $N_{max}$& $6$, $256$ &  $A$ & $1.2$ \\
$P_{B}$, $P_{s,B}$  & $20$\rm{dBm}, $30$\rm{dBm} & $P_U$, $P_n$& $10$\rm{dBm}\\
\hline
\end{tabular}
\label{T1}
\end{table*}

\section{Simulations}
\label{Sec:Sim}

Simulations results are obtained under the parameter values described in Table \ref{T1}.  Since the results are derived for a homogeneous network so we set the IRS-user path loss $\beta_2$ and BS-user path loss $\beta_d$ as the mean of the path loss of all users. We consider $5$ \rm{dBi} elements at the BS and IRS and penetration losses of $10$\rm{dB} for the IRS-assisted link and $25$\rm{dB} for the direct link.  The reflection coefficient $\alpha_n=1$, $\forall n$, as assumed in almost all works on IRS-assisted systems, motivated by the significant advancements made in the design of lossless metasurfaces \cite{MS_loss2}.

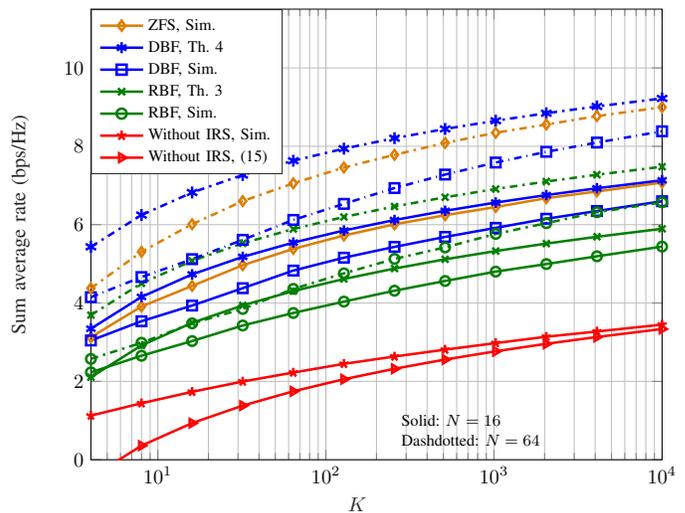
\begin{figure}[!t]
\centering
\tikzset{every picture/.style={scale=.95}, every node/.style={scale=.8}}
%
%
\definecolor{mycolor1}{rgb}{0.00000,0.49804,0.00000}%
\definecolor{mycolor2}{rgb}{0.87059,0.49020,0.00000}%
\begin{tikzpicture}


\begin{axis}[%
width=.95\columnwidth,
height=.75\columnwidth,
scale only axis,
xmode=log,
xmin=4,
xmax=10000,
xminorticks=true,
xlabel style={font=\color{white!15!black}},
xlabel={$K$},
ymin=0,
ymax=11.5,
ylabel style={at={(axis cs: 4,6)},font=\color{white!15!black}},
ylabel={Sum average rate (bps/Hz)},
axis background/.style={fill=white},
xmajorgrids,
xminorgrids,
ymajorgrids,
legend style={at={(axis cs: 4,11.5)},anchor=north west,legend cell align=left,align=left,draw=white!15!black, /tikz/column 2/.style={
                column sep=5pt,
            }},]
						
\addplot [color=mycolor2, line width=1.0pt,mark size=2.0pt, mark=diamond, mark options={solid, mycolor2}]
  table[row sep=crcr]{%
4	3.12802236339793\\
8	3.91135070024522\\
16	4.4360926013126\\
32	4.96534782995045\\
64	5.38165988384373\\
128	5.72271946626556\\
256	6.00826349193291\\
512	6.24133195900589\\
1024	6.44909581494424\\
2048	6.67610744492024\\
4096	6.84722321449087\\
10000	7.07407503093563\\
};
\addlegendentry{\footnotesize ZFS, Sim.}

\addplot [color=blue, line width=1.0pt,mark size=2.2pt, mark=asterisk, mark options={solid, blue}]
  table[row sep=crcr]{%
4	3.34519704146258\\
8	4.1561272576789\\
16	4.73149140258246\\
32	5.17777850521088\\
64	5.54242161879879\\
128	5.85072297845331\\
256	6.11778576370236\\
512	6.35335183501513\\
1024	6.56407286633078\\
2048	6.75469322593943\\
4096	6.92871597991869\\
10000	7.13258213424171\\
};
\addlegendentry{\footnotesize DBF, Th. \ref{Corr2}}

\addplot [color=blue, line width=1.0pt,mark size=2.0pt, mark=square, mark options={solid, blue}]
  table[row sep=crcr]{%
4	3.04360030399091\\
8	3.53564688314703\\
16	3.93532709488856\\
32	4.37880433842431\\
64	4.82782377663665\\
128	5.15695425495058\\
256	5.43269803001267\\
512	5.68720827985466\\
1024	5.91813359654609\\
2048	6.15059755062446\\
4096	6.34829990610066\\
10000	6.5977384171495\\
};
\addlegendentry{\footnotesize DBF, Sim.}

\addplot [color=mycolor1, line width=1.0pt,mark size=2.0pt, mark=x, mark options={solid, mycolor1}]
  table[row sep=crcr]{%
4	2.10757360722729\\
8	2.91850382344361\\
16	3.49386796834717\\
32	3.94015507097559\\
64	4.3047981845635\\
128	4.61309954421802\\
256	4.88016232946707\\
512	5.11572840077984\\
1024	5.32644943209549\\
2048	5.51706979170414\\
4096	5.6910925456834\\
10000	5.89495870000642\\
};
\addlegendentry{\footnotesize RBF, Th. \ref{Cor1}}

\addplot [color=mycolor1, line width=1.0pt,mark size=2.0pt, mark=o, mark options={solid, mycolor1}]
  table[row sep=crcr]{%
4	2.23161304055908\\
8	2.65482316645648\\
16	3.02933461871804\\
32	3.4260346035687\\
64	3.74578049759995\\
128	4.03640571517586\\
256	4.31446422289792\\
512	4.56274189364521\\
1024	4.80172123187384\\
2048	4.9944754223841\\
4096	5.19312593061832\\
10000	5.44037445631802\\
};
\addlegendentry{\footnotesize RBF, Sim.}

\addplot [color=red, line width=1.0pt,mark size=2.0pt, mark=star, mark options={solid, red}]
  table[row sep=crcr]{%
4	1.1264769653281\\
8	1.4414011345076\\
16	1.73339197386965\\
32	1.99582758055056\\
64	2.22504555469592\\
128	2.44753596870431\\
256	2.63437962721881\\
512	2.81001471318185\\
1024	2.97709411882554\\
2048	3.13973051336827\\
4096	3.27321428860912\\
10000	3.44931188198118\\
};
\addlegendentry{\footnotesize Without IRS, Sim.}

\addplot [color=red, line width=1.0pt,mark size=2.0pt, mark=triangle, mark options={solid, rotate=270, red}]
  table[row sep=crcr]{%
4	-0.450757123742257\\
8	0.360173092474071\\
16	0.935537237377631\\
32	1.38182434000605\\
64	1.74646745359396\\
128	2.05476881324848\\
256	2.32183159849752\\
512	2.55739766981029\\
1024	2.76811870112594\\
2048	2.95873906073459\\
4096	3.13276181471385\\
10000	3.33662796903687\\
};
\addlegendentry{\footnotesize Without IRS, \eqref{RBF_sc}}

\addplot [color=mycolor1, line width=1.0pt,mark size=2.0pt, dashdotted, mark=o, mark options={solid, mycolor1}]
  table[row sep=crcr]{%
4	2.57342761102706\\
8	2.98493510543829\\
16	3.48021543081358\\
32	3.84911011699586\\
64	4.36038003735884\\
128	4.75793472296732\\
256	5.12612235354212\\
512	5.41983417629702\\
1024	5.76384475782903\\
2048	6.03837087081956\\
4096	6.31793702778118\\
10000	6.5697396028389\\
};

\addplot [color=mycolor1, line width=1.0pt,mark size=2.0pt, dashdotted, mark=x, mark options={solid, mycolor1}]
  table[row sep=crcr]{%
4	3.69110689041342\\
8	4.50203710662976\\
16	5.07740125153332\\
32	5.52368835416173\\
64	5.88833146774964\\
128	6.19663282740417\\
256	6.4636956126532\\
512	6.69926168396597\\
1024	6.90998271528163\\
2048	7.10060307489028\\
4096	7.27462582886954\\
10000	7.47849198319255\\
};

\addplot [color=blue, dashdotted, line width=1.0pt,mark size=2.0pt, mark=square, mark options={solid, blue}]
  table[row sep=crcr]{%
4	4.14918704012632\\
8	4.66228051379873\\
16	5.11937512162496\\
32	5.61284325767875\\
64	6.1211605056732\\
128	6.53008186098865\\
256	6.9343386970914\\
512	7.27887524692098\\
1024	7.58573657456642\\
2048	7.85913786143944\\
4096	8.09561074566007\\
10000	8.38291026437167\\
};

\addplot [color=blue, line width=1.0pt,mark size=2.5pt, dashdotted, mark=asterisk, mark options={solid, blue}]
  table[row sep=crcr]{%
4	5.4343547417058\\
8	6.24528495792214\\
16	6.8206491028257\\
32	7.26693620545412\\
64	7.63157931904203\\
128	7.93988067869655\\
256	8.20694346394558\\
512	8.44250953525835\\
1024	8.65323056657401\\
2048	8.84385092618266\\
4096	9.01787368016192\\
10000	9.22173983448494\\
};

\addplot [color=mycolor2, line width=1.0pt,mark size=2.0pt, line width=1.0pt,mark size=2.0pt, dashdotted, mark=diamond, mark options={solid, mycolor2}]
  table[row sep=crcr]{%
4	4.37577692006199\\
8	5.30635029132634\\
16	6.01099170370773\\
32	6.60149752746534\\
64	7.05481198637271\\
128	7.45893792127624\\
256	7.78210486854164\\
512	8.08661140482841\\
1024	8.34575226853248\\
2048	8.55608151804133\\
4096	8.77051213218348\\
10000	8.9978315663506\\
};
%
%
\node at (axis cs: 256,1) [anchor = west] {\footnotesize Solid: $N=16$};
\node at (axis cs: 256,0.5) [anchor = west] {\footnotesize Dashdotted: $N=64$};

\end{axis}

\end{tikzpicture}%
\caption{Sum average rate for $M=2$. Simulated (Sim.) and theorem (Th.) results plotted.}
\label{Fig1}
\end{figure}

\begin{figure}[!t]
\centering
\tikzset{every picture/.style={scale=.95}, every node/.style={scale=.8}}
%
%
\definecolor{mycolor1}{rgb}{0.00000,0.49804,0.00000}%
\definecolor{mycolor2}{rgb}{0.87059,0.49020,0.00000}%
\begin{tikzpicture}

\begin{axis}[%
width=.95\columnwidth,
height=.75\columnwidth,
scale only axis,
xmode=log,
xmin=4,
xmax=10000,
xminorticks=true,
xlabel style={font=\color{white!15!black}},
xlabel={$K$},
ymin=0,
ymax=13,
ylabel style={at={(axis cs: 4.6,7)},font=\color{white!15!black}},
ylabel={Sum average rate (bps/Hz)},
axis background/.style={fill=white},
xmajorgrids,
xminorgrids,
ymajorgrids,
legend style={at={(axis cs: 4,13)},anchor=north west,legend cell align=left,align=left,draw=white!15!black, /tikz/column 2/.style={
                column sep=5pt,
            }},]

\addplot [color=mycolor2, line width=1.0pt,mark size=2.0pt, mark=diamond, mark options={solid, mycolor2}]
  table[row sep=crcr]{%
4	1.83666023913428\\
8	3.72505691870562\\
16	4.83823476783449\\
32	5.73768235398892\\
64	6.51692700961855\\
128	7.29802847038612\\
256	7.90637078111934\\
512	8.46490728420503\\
1024	9.00558008237942\\
2048	9.49907094969097\\
4096	9.93150071601236\\
10000	10.340398348177\\
};
\addlegendentry{\footnotesize ZFS, Sim.}

\addplot [color=blue, line width=1.0pt,mark size=2.0pt, mark=asterisk, mark options={solid, blue}]
  table[row sep=crcr]{%
4	3.526115536926\\
8	5.14797596935865\\
16	6.29870425916579\\
32	7.19127846442262\\
64	7.92056469159844\\
128	8.53716741090747\\
256	9.07129298140556\\
512	9.5424251240311\\
1024	9.9638671866624\\
2048	10.3451079058797\\
4096	10.6931534138382\\
10000	11.1008857224843\\
};
\addlegendentry{\footnotesize DBF, Th. \ref{Corr2}}

\addplot [color=blue, line width=1.0pt,mark size=2.0pt, mark=square, mark options={solid, blue}]
  table[row sep=crcr]{%
4	3.31385725469554\\
8	3.87829847358524\\
16	4.55501610587125\\
32	5.13009364379684\\
64	5.62441145255081\\
128	6.08203276044668\\
256	6.54334262435861\\
512	7.00154254081031\\
1024	7.55582234983187\\
2048	8.06561599674869\\
4096	8.45349714806591\\
10000	8.91855024868751\\
};
\addlegendentry{\footnotesize DBF, Sim.}

\addplot [color=mycolor1, line width=1.0pt,mark size=2.0pt, mark=x, mark options={solid, mycolor1}]
  table[row sep=crcr]{%
4	-0.421109897269901\\
8	1.20075053516275\\
16	2.35147882496989\\
32	3.24405303022672\\
64	3.97333925740254\\
128	4.58994197671157\\
256	5.12406754720966\\
512	5.5951996898352\\
1024	6.0166417524665\\
2048	6.3978824716838\\
4096	6.70592797964232\\
10000	7.05366028828836\\
};
\addlegendentry{\footnotesize RBF, Th. \ref{Cor1}}

\addplot [color=mycolor1, line width=1.0pt,mark size=2.0pt, mark=o, mark options={solid, mycolor1}]
  table[row sep=crcr]{%
4	1.49433454903223\\
8	1.92572026162172\\
16	2.33152081122197\\
32	2.77745234806296\\
64	3.18180594609861\\
128	3.54267154765953\\
256	3.88758781916768\\
512	4.23559475345564\\
1024	4.52975319834924\\
2048	4.82627902372403\\
4096	5.09934812251982\\
10000	5.43271079078426\\
};
\addlegendentry{\footnotesize RBF, Sim.}

\addplot [color=red, line width=1.0pt,mark size=2.0pt, mark=star, mark options={solid, red}]
  table[row sep=crcr]{%
4	0.933493907509083\\
8	1.38000668207188\\
16	1.77965182495724\\
32	2.12003527468366\\
64	2.40780817324108\\
128	2.69554368074443\\
256	2.99287494276265\\
512	3.22725564435593\\
1024	3.48523656218419\\
2048	3.72834978301699\\
4096	3.95886361879787\\
10000	4.24103834808214\\
};
\addlegendentry{\footnotesize Without IRS, Sim.}

\addplot [color=red, line width=1.0pt,mark size=2.0pt, mark=triangle, mark options={solid, rotate=270, red}]
  table[row sep=crcr]{%
4	-3.67410296972429\\
8	-2.05224253729163\\
16	-0.901514247484513\\
32	-0.00894004222767819\\
64	0.720346184948141\\
128	1.33694890425718\\
256	1.87107447475528\\
512	2.3422066173808\\
1024	2.76364868001211\\
2048	3.14488939922941\\
4096	3.49293490718793\\
10000	3.90066721583396\\
};
\addlegendentry{\footnotesize Without IRS, \eqref{RBF_sc}}

\end{axis}
\end{tikzpicture}%
\caption{Sum average rate  for $M=4$, $N=32$. Both Sim. and Th. results are plotted.}
\label{Fig2}
\end{figure}
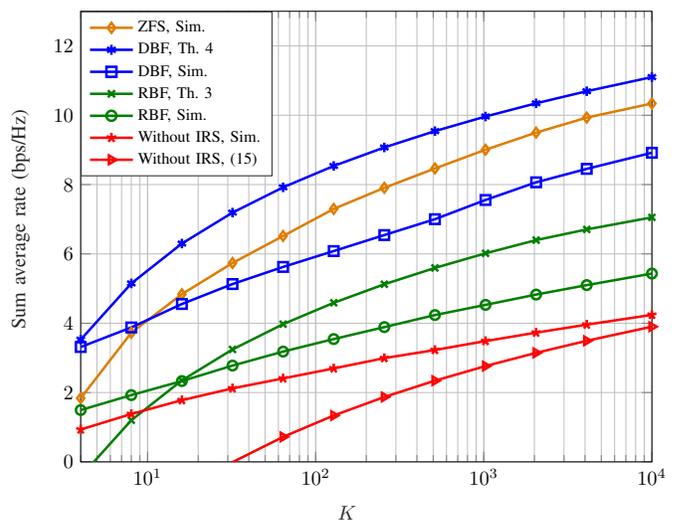

We first study in Fig. \ref{Fig1} the sum average rate performance against $K$ under RBF and DBF at the BS and the random rotations scheme at the IRS in a MISO BC. The sum-rate is seen to increase with $K$ due to the multi-user diversity effect, i.e. as the number of users increases, it becomes more likely to have some users close to their channel peaks. By scheduling these users, the sum average rate  increases with $K$. The simulated sum-rate plotted using \eqref{exp} is shown to scale according to Theorem \ref{Cor1} for RBF and Theorem \ref{Corr2} for DBF as $K$ grows large (i.e. have approximately the same slope). As seen in the derived scaling laws, the random rotations IRS scheme provides a sum-rate gain of $M\log \left(N \alpha^2 \frac{\beta_r}{\beta_d}+1\right)$ over the conventional system but also causes a rate loss of approximately $\log \det \bar{\mathbf{R}}$ due to the correlation introduced by $\mathbf{H}_1$. Fig. \ref{Fig1} shows that the positive effect of the improved array gain outweighs the rate loss caused by the correlation resulting in a significantly better performance than the system without IRS. In Fig. \ref{Fig1}, we see that introducing an IRS yields approximately a $3.3$ and $5$ bit/transmission improvement with $N=16$ and $64$ elements respectively under DBF as compared to the system without IRS.

We also show the performance of ZFS \cite{ZF_GUS} discussed in Sec. III-D, where instead of sending random beams to the strongest users, users are selected on the basis of the full CSI and zero-forcing is employed at the BS. The IRS still continues to employ random phase rotations.  The ZFS scheme is shown to achieve a  significant portion of the average sum-capacity promised by DPC in \cite{ZF_GUS} and is therefore considered here for comparison, since online implementation of DPC based schemes is computationally prohibitive for large $K$. We observe that the sum average rate scaling (i.e. the slope of the curves) under this scheme and DBF are approximately the same as $K$ becomes large, confirming that DBF (without requiring instantaneous CSI) asymptotically achieves the same scaling as ZFS that requires full CSI from all users.

In Fig. \ref{Fig2}, we plot the same results for $M=4$ and $N=32$. The sum average rate increases with  $M$ since a higher number of users are simultaneously served but the gain is not linear  due to the decrease in the SNR term $\frac{P_T}{M \sigma^2}$ with $M$. The convergence of the slope of the simulated sum average rate to the scaling law slows down for large $M$, with a higher $K$ needed to obey the scaling laws. This is in accordance with the result in \cite{RBF} that the number of users should grow exponentially in $M$ to obey the scaling laws. However, 5G networks target to provide massive connectivity making the studied regime relevant.  Here we also remark that the scaling laws in Theorem \ref{Cor1} and Theorem \ref{Corr2} describe how the sum average rate in \eqref{exp} ``scales" for large $K$ and are not to be considered as its approximations.

\begin{figure}[!t]
\centering
\tikzset{every picture/.style={scale=.95}, every node/.style={scale=.8}}
%
%
\definecolor{mycolor1}{rgb}{0.00000,0.49804,0.00000}%
\definecolor{mycolor2}{rgb}{1.00000,0.00000,1.00000}%
\begin{tikzpicture}

\begin{axis}[%
width=.95\columnwidth,
height=.75\columnwidth,
scale only axis,
xmode=log,
xmin=4,
xmax=10000,
xminorticks=true,
xlabel style={font=\color{white!15!black}},
xlabel={$K$},
ymin=0,
ymax=6.8,
ylabel style={at={(axis cs: 4.5,3.5)},font=\color{white!15!black}},
ylabel={Sum average rate (bps/Hz)},
axis background/.style={fill=white},
xmajorgrids,
xminorgrids,
ymajorgrids,
legend style={at={(axis cs: 10000,0)},anchor=south east,legend cell align=left,align=left,draw=white!15!black, /tikz/column 2/.style={
                column sep=5pt,
            }},]
\addplot [color=black, line width=1.0pt,mark size=2.0pt, mark=o, mark options={solid, black}]
  table[row sep=crcr]{%
4	4.25479323280745\\
8	4.65908583833301\\
16	4.94485168020679\\
32	5.1735570781316\\
64	5.42089085204682\\
128	5.59437151589527\\
256	5.78818384874123\\
512	5.9178339861305\\
1024	6.05906342093145\\
2048	6.18005199025764\\
4096	6.28961568761874\\
10000	6.40857383\\
};
\addlegendentry{\footnotesize DBF, Coh. BF ($b=3$)}

\addplot [color=black, line width=1.0pt,mark size=2.0pt,mark=diamond, mark options={solid, black}]
  table[row sep=crcr]{%
4	3.94379722845195\\
8	4.23826456616034\\
16	4.56561258445768\\
32	4.81647016551032\\
64	5.0463079231634\\
128	5.23439936481738\\
256	5.40815718181563\\
512	5.57270890019321\\
1024	5.72469629320991\\
2048	5.86721251548137\\
4096	5.98767953265348\\
10000	6.11075622560592\\
};
\addlegendentry{\footnotesize DBF, Coh. BF ($b=2$)}

\addplot [color=blue, line width=1.0pt,mark size=2.2pt, mark=asterisk, mark options={solid, blue}]
  table[row sep=crcr]{%
4	2.01667658915795\\
8	2.82760680537428\\
16	3.40297095027784\\
32	3.84925805290626\\
64	4.21390116649417\\
128	4.52220252614869\\
256	4.78926531139773\\
512	5.0248313827105\\
1024	5.23555241402615\\
2048	5.4261727736348\\
4096	5.60019552761406\\
10000	5.80406168193709\\
};
\addlegendentry{\footnotesize DBF, Rand., Th. \ref{Corr2}}

\addplot [color=blue, line width=1.0pt,mark size=2.0pt, dashed, mark=+, mark options={solid, blue}]
  table[row sep=crcr]{%
4	2.62758057180841\\
8	3.08936369706703\\
16	3.42791868776539\\
32	3.7633253589802\\
64	4.03491683391329\\
128	4.32355643788068\\
256	4.59580896168794\\
512	4.84797515364052\\
1024	5.04227539792346\\
2048	5.24326534224562\\
4096	5.42371422570065\\
10000	5.644838292\\
};
\addlegendentry{\footnotesize DBF, Rand. (Continuous), Sim.}

\addplot [color=blue, line width=1.0pt,mark size=2.0pt, mark=square, mark options={solid, blue}]
  table[row sep=crcr]{%
4	2.60213170238388\\
8	3.07359968176468\\
16	3.39760302489789\\
32	3.73383754993192\\
64	4.01915050429042\\
128	4.31884743401896\\
256	4.57809143113696\\
512	4.82231181582197\\
1024	5.01528939743623\\
2048	5.22237961824822\\
4096	5.41923774674092\\
10000	5.638372822\\
};
\addlegendentry{\footnotesize DBF, Rand. ($b=2$), Sim.}

\addplot [color=red, line width=1.0pt,mark size=2.0pt, mark=star, mark options={solid, red}]
  table[row sep=crcr]{%
4	1.14808985868062\\
8	1.53351656186974\\
16	1.78881870184456\\
32	2.02707159101719\\
64	2.25949609213954\\
128	2.45540307130589\\
256	2.63878821338798\\
512	2.843399237452\\
1024	2.99965576610446\\
2048	3.16076777599404\\
4096	3.29922450340687\\
10000	3.4728685\\
};
\addlegendentry{\footnotesize Without IRS, Sim.}

\addplot [color=red, line width=1.0pt,mark size=2.0pt, mark=triangle, mark options={solid, rotate=270, red}]
  table[row sep=crcr]{%
4	-0.450757123742257\\
8	0.360173092474071\\
16	0.935537237377631\\
32	1.38182434000605\\
64	1.74646745359396\\
128	2.05476881324848\\
256	2.32183159849752\\
512	2.55739766981029\\
1024	2.76811870112594\\
2048	2.95873906073459\\
4096	3.13276181471385\\
10000	3.33662796903687\\
};
\addlegendentry{\footnotesize Without IRS, \eqref{RBF_sc}}

\end{axis}
\end{tikzpicture}%
\caption{Performance under random (Rand.) rotations and coherent beamforming  (Coh. BF).}
\label{Fig3}
\end{figure}
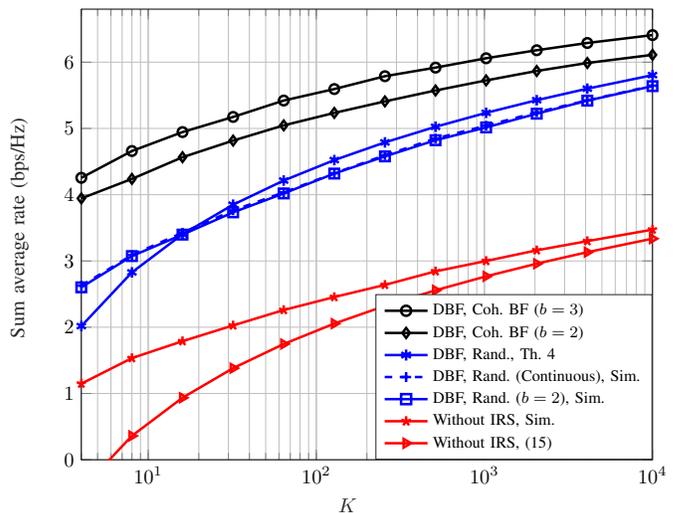

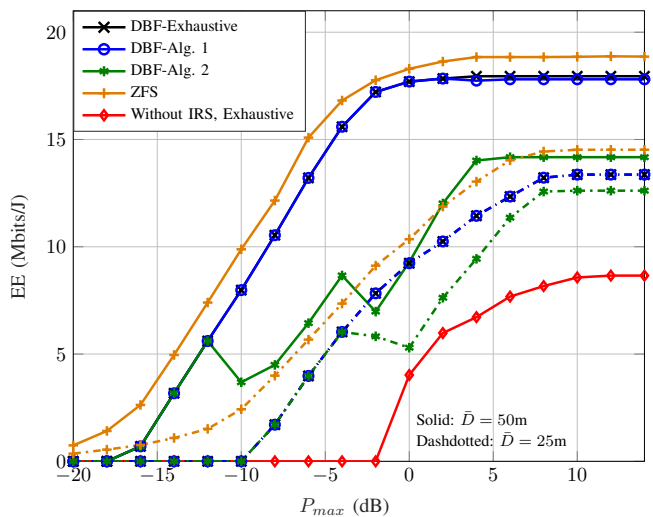
\begin{figure}[!t]
\centering
\tikzset{every picture/.style={scale=.95}, every node/.style={scale=.8}}
%
%
\definecolor{mycolor1}{rgb}{0.87059,0.49020,0.00000}%
\definecolor{mycolor2}{rgb}{0.00000,0.49804,0.00000}%
\definecolor{mycolor3}{rgb}{0.74902,0.00000,0.74902}%

\begin{tikzpicture}

\begin{axis}[%
width=.95\columnwidth,
height=.75\columnwidth,
scale only axis,
xmin=-20,
xmax=14,
xlabel style={font=\color{white!15!black}},
xlabel={$P_{max}$ (\rm{dB})},
ymin=0,
ymax=21,
ylabel style={at={(axis cs: -19,10)},font=\color{white!15!black}},
ylabel={EE (Mbits/J)},
axis background/.style={fill=white},
xmajorgrids,
ymajorgrids,
legend style={at={(axis cs: -20,21)},anchor=north west,legend cell align=left,align=left,draw=white!15!black, /tikz/column 2/.style={
                column sep=5pt,
            }},]

\addplot [color=black,  line width=1.0pt,mark size=3pt, mark=x, mark options={solid, black}]
  table[row sep=crcr]{%
-20	0\\
-18	0\\
-16	0.697821102248104\\
-14	3.1718930736526\\
-12	5.60713215577236\\
-10	7.98046147333004\\
-8	10.5391933174419\\
-6	13.2093921518029\\
-4	15.5937809263598\\
-2	17.2145674147557\\
0	17.6976734187633\\
2	17.8385992921117\\
4	17.9476546170602\\
6	17.9476546170602\\
8	17.9476546170602\\
10	17.9476546170602\\
12	17.9476546170602\\
14	17.9476546170602\\
};
\addlegendentry{\footnotesize DBF-Exhaustive}

\addplot [color=blue,  line width=1.0pt,mark size=2.0pt, mark=o, mark options={solid, blue}]
  table[row sep=crcr]{%
-20	0\\
-18	0\\
-16	0.697821102248104\\
-14	3.1718930736526\\
-12	5.60713215577235\\
-10	7.98046147333005\\
-8	10.5391933174419\\
-6	13.2093921518029\\
-4	15.5937809263598\\
-2	17.2145674147557\\
0	17.6976734187633\\
2	17.8385992921117\\
4	17.7413680094699\\
6	17.8040205877366\\
8	17.8040205877366\\
10	17.8040205877366\\
12	17.8040205877366\\
14	17.8040205877366\\
};
\addlegendentry{\footnotesize DBF-Alg. 1}

\addplot [color=mycolor2,  line width=1.0pt,mark size=2.0pt, mark=asterisk, mark options={solid, mycolor2}]
  table[row sep=crcr]{%
-20	0\\
-18	0\\
-16	0.697821102248104\\
-14	3.1718930736526\\
-12	5.60713215577235\\
-10	3.685844\\
-8	4.49369573195256\\
-6	6.44204825767576\\
-4	8.66190372705337\\
-2	6.9786552432614\\
0	9.28133452361284\\
2	12.0179278971202\\
4	14.020136382085\\
6	14.1750136382085\\
8	14.1750136382085\\
10	14.1750136382085\\
12	14.1750136382085\\
14	14.1750136382085\\
};
\addlegendentry{\footnotesize DBF-Alg. 2}

\addplot [color=mycolor1, mark=+,  line width=1.0pt,mark size=2.0pt, mark options={solid, mycolor1}]
  table[row sep=crcr]{%
-20	0.74283829222\\
-18	1.414534894196625\\
-16	2.62773697185672\\
-14	4.95827067075154\\
-12	7.40069929606078\\
-10	9.88265165622267\\
-8	12.1540110559667\\
-6	15.0874564174846\\
-4	16.8165104887595\\
-2	17.7624422598419\\
0	18.2945518879956\\
2	18.6336285349431\\
4	18.8406539505024\\
6	18.8409332457417\\
8	18.8414046278274\\
10	18.8534196437164\\
12	18.8758700006025\\
14	18.8632002426159\\
};
\addlegendentry{\footnotesize ZFS}

\addplot [color=red, line width=1.0pt,mark size=2.0pt,  mark=diamond, mark options={solid, red}]
  table[row sep=crcr]{%
-20	0\\
-18	0\\
-16	0\\
-14	0\\
-12	0\\
-10	0\\
-8	0\\
-6	0\\
-4	0\\
-2	0\\
0	4.02227337747481\\
2	5.97964043440897\\
4	6.7195294743187\\
6	7.67730134016267\\
8	8.16735864334166\\
10	8.56865182018446\\
12	8.65585266225369\\
14	8.65585266225369\\
};
\addlegendentry{\footnotesize Without IRS, Exhaustive}

\addplot [color=black, dashdotted, line width=1.0pt,mark size=3.0pt, mark=x, mark options={solid, black}]
  table[row sep=crcr]{%
-20	0\\
-18	0\\
-16	0\\
-14	0\\
-12	0\\
-10	0\\
-8	1.7080541036279\\
-6	3.9793634456192\\
-4	6.03079219664502\\
-2	7.82954097189912\\
0	9.23293090155856\\
2	10.2476471464112\\
4	11.4416796398562\\
6	12.3407722387525\\
8	13.2185460801836\\
10	13.3623680539963\\
12	13.3623680539963\\
14	13.3623680539963\\
};

\addplot [color=blue, dashdotted, line width=1.0pt,mark size=2.0pt, mark=o, mark options={solid, blue}]
  table[row sep=crcr]{%
-20	0\\
-18	0\\
-16	0\\
-14	0\\
-12	0\\
-10	0\\
-8	1.7080541036279\\
-6	3.97936344561921\\
-4	6.03079219664502\\
-2	7.82954097189911\\
0	9.23293090155855\\
2	10.2476471464112\\
4	11.4416796398562\\
6	12.3407722387525\\
8	13.2185460801836\\
10	13.3623901111049\\
12	13.3623901111049\\
14	13.3623901111049\\
};

\addplot [color=mycolor2,  line width=1.0pt,dashdotted,mark size=2.0pt, mark=asterisk, mark options={solid, mycolor2}]
  table[row sep=crcr]{%
-20	0\\
-18	0\\
-16	0\\
-14	0\\
-12	0\\
-10	0\\
-8	1.7080541036279\\
-6	3.97936344561921\\
-4	6.03079219664502\\
-2	5.82494284372458\\
0	5.29614739949648\\
2	7.62345708304942\\
4	9.43682963955456\\
6	11.3567990710846\\
8	12.574104086773\\
10	12.614104086773\\
12	12.614104086773\\
14	12.614104086773\\
};

\addplot [color=mycolor1,line width=1.0pt,mark size=2.0pt,  dashdotted, mark=+, mark options={solid, mycolor1}]
  table[row sep=crcr]{%
-20	0.360008704581814\\
-18	0.54556278027473\\
-16	0.757150653453474\\
-14	1.089169769195439\\
-12	1.51289955279754\\
-10	2.42817128321463\\
-8	3.99389688493044\\
-6	5.67362814522528\\
-4	7.35242498239076\\
-2	9.11835631528389\\
0	10.3527470138999\\
2	11.8731106733849\\
4	13.0376830020314\\
6	14.0311533673719\\
8	14.4406515941279\\
10	14.5206958910625\\
12	14.5232404801805\\
14	14.5230023886385\\
};

\node at (axis cs: 0,2) [anchor = west] {\footnotesize Solid: $\bar{D}=50$m};
\node at (axis cs: 0,1) [anchor = west] {\footnotesize Dashdotted: $\bar{D}=25$m};

\end{axis}
\end{tikzpicture}%
\caption{EE against $P_{max}$ for the IRS-assisted BC under random rotations IRS scheme.}
\label{Fig5}
\end{figure}

Next, we discuss how the performance of the proposed random rotations IRS scheme compares to coherent beamforming,  where the IRS phase shifts are optimized based on instantaneous CSI. The coherent IRS beamforming is implemented using exhaustive search for $N=6$ as follows. The BS obtains perfect CSI of the individual  links (i.e. $\mathbf{h}_{2,k}$'s and $\mathbf{h}_{d,k}$'s) from all users, which requires prohibitively long training time under the estimation protocols in \cite{LS1} and \cite{annie_OJ}.  It then calculates $\underset{k=1,\dots, K}{\text{max}}\gamma_{k,m}$ in \eqref{exp} for every possible combination of phase-shifts implementable using $b$-bit IRS phase shifters. That is, each IRS element can employ a discrete phase shift chosen from the set $\mathcal{F}=\{0, \bigtriangleup \theta, \dots, (Q-1)	\bigtriangleup \theta \}$ where $\bigtriangleup \theta=\frac{2\pi}{Q}$ and $Q=2^b$. The sum average rate  in \eqref{exp} is computed for all possible combinations of IRS phase shifts and the combination ($\theta_1 \in \mathcal{F}, \dots, \theta_N \in \mathcal{F}$) that maximizes the sum rate is adopted in that coherence interval. Therefore, this serves as a performance upper-bound for IRS-assisted MISO BC. The performance of random rotations IRS scheme, in which the IRS elements apply random phase rotations without requiring any CSI  while the BS implements DBF using partial SINR feedback from users as explained in Sec. III-B, is also plotted. For fairness in comparison, we draw the phase shifts for the random rotations scheme randomly from the same discrete set $\mathcal{F}$ used for exhaustive search. We also plot the performance of the random rotations scheme under the scenario where the IRS phase shifts are selected randomly from the continuous range $[0,2\pi]$.

\begin{table*}[!b]
\caption{EE performance under different algorithms.}
\centering
    \begin{tabular}{|c|c|c|c|c|c|c|c|c|}
    \hline
    &\multicolumn{4}{|c|}{$\bar{\textbf{D}}\textbf{=50m}$}& \multicolumn{4}{|c|}{$\bar{\textbf{D}}\textbf{=25m}$} \\ \hline
		& $\textbf{M}^*$ & $\textbf{N}^*$ &$\textbf{P}_T^*$ & \textbf{Optimal $\text{EE}^{(K)}$ value} & $\textbf{M}^*$ & $\textbf{N}^*$ &$\textbf{P}_T^*$ & \textbf{Optimal $\text{EE}^{(K)}$ value} \\ \hline
    \textbf{Exhaustive} & 2& 116& 1.9& 17.94& 6 & 224 & 7.48 & 13.36 \\ \hline
    \textbf{Algorithm 1}&3 &140 &2.8 & 17.80& 6 & 224& 7.48 & 13.36\\ \hline
    \textbf{Algorithm 2}&5&212&1.86 &14.17 & 6 & 256 & 4.25 & 12.61 \\ \hline
    \textbf{Without IRS}& 6&-&10&8.57 &6&-&10&8.57 \\ \hline
    \end{tabular}
\label{T2}
 \end{table*}

Interestingly, we observe in Fig. \ref{Fig3} that the sum average rate gap between the random rotations scheme and coherent beamforming decreases with $K$.  This is because as $K$ increases, it becomes more likely that  even under the random rotations scheme there are $M$ users in the network who are in their respective coherent beamforming configurations with respect to the random phase shifts employed at the IRS, i.e. their channels have the optimal relationship with the IRS phase-shifts adopted in that coherence interval. By scheduling these $M$ strong users, the performance under random IRS phase shifts starts getting closer to the scenario where IRS phase shifts are chosen using exhaustive search based on full CSI.  This makes our scheme very desirable in the large user regime especially under opportunistic scheduling (i.e. RBF, DBF). Also note that while the performance of coherent beamforming improves with the resolution of IRS phase shifters $b$, the performance of the random rotations scheme is insensitive to whether the phase shifts at the IRS are chosen from a discrete set or a continuous set. This is because our scheme does not exploit any reflect beamforming gains at the IRS but rather just uses the IRS to increase the multi-user diversity effect by increasing the variance of the channel elements, defined in \eqref{R_def}, by approximately a factor of $N\frac{\beta_r}{\beta_d}+1$. This increase in the range of channel fluctuations (which determines the multi-user diversity gain) is independent of the choice of phase shifts at the IRS.

Next we study the EE performance of the IRS-assisted MISO BC under DBF at the BS and the random rotations scheme at the IRS. The EE scaling is given  by \eqref{EE}, with parameters defined in Table \ref{T1}. The optimal values of the number of BS antennas $M^*$, number of IRS elements $N^*$ and transmit power $P_T^*$ are found using (i) exhaustive search outlined in Section \ref{Sec:Opt}-B with $\delta=0.01$, (ii) Algorithm \ref{Alg:ExactAM} which solves the exact problem in (\textit{P1}), and (iii) Algorithm \ref{Alg:ApproxAM} which solves an upper bound on the exact problem specified in (\textit{P2}), for which we tabulate the actual EE in \eqref{EE} (not the upper-bound value). The performance without IRS is also evaluated where $M^*$ and $P_T^*$ are found using exhaustive search. The results are tabulated in Table \ref{T2}. The performance yielded by Algorithm \ref{Alg:ExactAM} is very close to the global solution obtained using exhaustive search with the former yielding an EE of $17.80$\rm{Mbits/J} and the latter yielding $17.94$\rm{Mbits/J}. The performance under Algorithm \ref{Alg:ApproxAM} yields a  lower EE value of around $14.17$\rm{Mbits/J} with higher  numbers of BS antennas and IRS elements needed to be activated.  However, as the BS-IRS distance is reduced to $25\rm{m}$,  the bound in \eqref{bounbd} becomes closer to the true EE in \eqref{EE} and the solution of Algorithm \ref{Alg:ApproxAM} (i.e. $12.61$\rm{Mbits/J}) becomes closer to the global solution (i.e. $13.36$\rm{Mbits/J}). 

Here we remark that while Algorithm 2 leads to higher values of $M$ and $N$ that need to be activated, this does not necessarily lead  to a higher hardware complexity. The considered IRS-assisted system is built with $M_{max}$ antennas at the BS and $N_{max}$ reflecting elements at the IRS, and  a subset of these are activated as a result of solving the EE maximization problem. So, in any case, we need to build the same hardware. While Algorithm 2 does not generally perform as well as the exact algorithms, the performance gap does significantly reduce for small values of $\bar{D}$, for example for $\bar{D}=25m$, with the advantage of a significantly reduced computational complexity.  Therefore, Algorithm 2 is a useful alternate method to optimize system parameters when $\bar{D}$ has small to moderate values and computational complexity is a critical issue.

These numbers are also illustrated in Fig. \ref{Fig5}, where we plot $\text{EE}^{(K)}$ in \eqref{EE} against $P_{max}$ for the optimal values $M^*$, $N^*$ and $P_T^*$ computed using exhaustive search, Algorithm 1 and Algorithm 2 under DBF at the BS and random rotations scheme at the IRS. The maximum value taken by EE under each optimization method at $P_{max}=10$\rm{dB} matches the values stated in Table \ref{T2}. The performance under all three methods is much better than the system without the IRS. We also simulate and plot the EE performance under ZFS scheme at the BS for values of $M^*$, $N^*$ and $P_T^*$ obtained using exhaustive search under DBF. Note that the sum rate scaling of ZFS is not known so we can not optimize the system parameters directly for ZFS under our optimization framework. The EE under ZFS is better than that under DBF, since ZFS achieves higher sum average rate (at the cost of higher feedback overhead and complexity as outlined in Table I). 

Fig. \ref{Fig5} also shows that the performance gap between Algorithm 1 and Algorithm 2 is noticeable for $\bar{D}=50$m. However, for $\bar{D}=25$m, Algorithm 2 performs very close to Algorithm 1 and exhaustive search since the bound in \eqref{bounbd} over which we optimize in \textit{(P2)} becomes closer to the true $\text{EE}^{(K)}$ in \eqref{EE} that is being plotted. Also notice that there are some jumps in the $\text{EE}$ curve under Algorithm 2 for intermediate values of $P_{max}$, because of optimizing the upper bound. These jumps represent a change in optimal values of system parameters that increased the bound but not the true function being plotted. However, for $P_{max}>0$\rm{dB} the jumps settle down and the curves eventually saturate. To summarize, Algorithm 1 always performs very close to the global solution obtained using exhaustive search and promises large EE gains by using an IRS in the MISO BC. Algorithm 2 can be utilized as a low-complexity alternative when BS-IRS distance is not large (for example in small cell settings) and computational cost is an important factor. 

\section{Conclusion}
\label{Sec:Con}

This work makes the preliminary contribution of studying the IRS-enabled random rotations scheme in an IRS-assisted MISO BC, under which the reflecting elements only introduce  random phase rotations without requiring instantaneous CSI, making it desirable in terms of implementation as compared to coherent beamforming. Under this framework, we derive the scaling laws of the average sum-capacity (achieved by DPC under full CSI) as well as that of the sum average rate achieved by RBF and DBF schemes under partial CSI at the BS.  We show that the random rotations scheme increases the sum-rate by exploiting the multi-user diversity effect, but also compromises the gain to some extent due to the correlation in the IRS-assisted channel. The results are used to formulate the EE scaling law, which is maximized in terms of the number of BS antennas, IRS elements and transmit power. Simulations show the proposed IRS scheme to approach the coherent beamforming  performance for a large number of users.

\appendices

\section{Proof of Theorem \ref{Thm1}}
\label{App:DPC}

The IRS-assisted channel in \eqref{ch_RS_RB} under the random rotations scheme is statistically equivalent to  $\mathbf{h}_k=\mathbf{R}^{1/2} \mathbf{z}_k$ where $\mathbf{z}_k\sim \mathcal{CN}(\mathbf{0}, \mathbf{I}_M)$ \eqref{ch_corr_RBF}. The sum-capacity in \eqref{rdpc} can then be written as
\begin{align}
\label{DPC_sc_RS1}
&R_{sum}=\mathbb{E}\Big[\underset{\underset{\sum_{k=1}^K p_k = P_T}{p_1, \dots, p_K}}{\text{max}} \log \Big(\text{det} \left( \mathbf{R}^{-1}+\frac{1}{\sigma^2} \sum_{k=1}^K  p_k \mathbf{z}_k \mathbf{z}_k^H \right) \nonumber \\
& \text{det } \mathbf{R} \Big)  \Big].
\end{align}
Using $\text{det}\mathbf{A}\leq \left( \frac{\text{tr}(\mathbf{A})}{M} \right)^M$ and $\sum_{k=1}^K \text{tr}(p_k \mathbf{z}_k \mathbf{z}_k^H)  \leq \underset{k}{\text{max }} \text{tr} (\mathbf{z}_k \mathbf{z}_k^H)  \sum_{k=1}^K p_k= \underset{k}{\text{max }} ||\mathbf{z}_k||^2 P_T$ yields
\begin{align} 
\label{temp1}
&\log \text{det} \left( \mathbf{R}^{-1}+\frac{1}{\sigma^2} \sum_{k=1}^K p_k \mathbf{z}_k  \mathbf{z}_k^H \right) \leq  M \log\Big(\frac{1}{M} \text{tr} \mathbf{R}^{-1} \nonumber \\
&+ \underset{k}{\text{max }} ||\mathbf{z}_k ||^2 \frac{P_T }{M \sigma^2} \Big).
\end{align}
All $\mathbf{z}_k$'s are i.i.d., so we can apply Lemma \ref{Lemma1} to study the behaviour of $\underset{k}{\text{max}} ||\mathbf{z}_k ||^2$. We note that $Z=||\mathbf{z}_k||^2$ has $\chi^2(2M)$ distribution. Using the PDF and CDF of a $\chi^2(2M)$ RV, we can verify the condition in \eqref{growth} as $g(z)=\frac{1-F_Z(z)}{f_Z(z)}=\frac{(M-1)!\exp(-z) \sum_{i=0}^{M-1}\frac{z^i}{i!}}{\exp(-z) z^{M-1}}=(M-1)!\sum_{i=0}^{M-1} \frac{z^{i-(M-1)}}{i!} \geq 0$. Finally using $1-F_Z(l_K)=\frac{1}{K}$,  we can show that  $\underset{k}{\text{max}} ||\mathbf{z}_k ||^2$ behaves like $l_K=\log K + (M-1)\log\log K +O(\log \log \log K)$ as $K\rightarrow \infty$. Therefore for large $K$, we obtain
\begin{align}
\label{temp2}
&R_{sum}^{(K)} \leq M \log\left(\frac{1}{M} \text{tr} \mathbf{R}^{-1} + \frac{P_T}{M \sigma^2} \log K +O(\log \log K) \right) \nonumber \\
&+\log \text{det } \mathbf{R}+o(1).
\end{align}
Using the fact that for large $K$, $\log\left(\frac{1}{M} \text{tr} \mathbf{R}^{-1} + \frac{P_T}{M \sigma^2} \log K +O(\log \log K) \right)=\log\Big(\left(\frac{P_T}{M \sigma^2} \log K \right) \Big(\frac{\text{tr} \mathbf{R}^{-1}}{\frac{P_T}{\sigma^2}\log K} +1+ O\left(\frac{\log \log K}{\log K}\right) \Big)  \Big)=\log \frac{P_T}{M \sigma^2} + \log \log K + o(1)$, we can further simplify (\ref{temp2}) to
\begin{align}
\label{temp3}
&R_{sum}^{(K)} \leq M \log \frac{P_T}{M \sigma^2} + M \log \log K  + M \log (\text{det } \mathbf{R})^{1/M} + o(1).
\end{align}
Next we  define $\bar{\mathbf{R}}=\frac{1}{\beta_r \sum_{n=1}^N \alpha_n^2 +\beta_d} \mathbf{R}$ such that $\text{tr}(\bar{\mathbf{R}})=M$, and write \eqref{temp3} as
\begin{align}
\label{temp33}
&R_{sum}^{(K)} \leq M \log \frac{P_T}{M \sigma^2} + M \log ((\beta_r \sum_{n=1}^N \alpha_n^2 +\beta_d) \log K)  \nonumber \\
&+ M \log (\text{det } \bar{\mathbf{R}})^{1/M} + o(1),
\end{align}
for large $K$. This yields the desired upper bound to the average sum-capacity scaling.

While the average sum-capacity is achievable using DPC, we find that the asymptotic scaling in \eqref{temp33} is also achievable using a low-complexity DBF scheme studied later in Sec. \ref{Sec:Asym3}. Using Theorem \ref{Corr2}  from Sec. \ref{Sec:Asym3}, we get 
\begin{align}
\label{temp4}
&R_{sum}^{(K)} \geq M \log \frac{P_T}{\sigma^2 M} + M \log ((\beta_r \sum_{n=1}^N \alpha_n^2 +\beta_d) \log K)  \nonumber \\
&+ M \log (\text{det } \bar{\mathbf{R}})^{1/M} + o(1),
\end{align}
for large $K$, confirming the result in \eqref{temp33} is achievable. Using (\ref{temp33}) and (\ref{temp4}) yields Theorem \ref{Thm1}.

\section{Proof of Lemma \ref{LemmaLamb}}
\label{App:Lamb}
$\lambda_M$ is the maximum eigenvalue of $\mathbf{A}_m$, which can be written using the definition of $\mathbf{A}_m$ as
\begin{align}
&\lambda_M=\underset{||\mathbf{u}||_2=1}{\text{max}} \mathbf{u}^H \mathbf{A}_m \mathbf{u}=\underset{||\mathbf{u}||_2=1}{\text{max}}\mathbf{u}^H (\boldsymbol{\Lambda}^{1/2} \bar{\boldsymbol{\phi}}_m \bar{\boldsymbol{\phi}}_m^H \boldsymbol{\Lambda}^{1/2}\nonumber \\
&-x \boldsymbol{\Lambda}^{1/2} \sum_{i\neq m} \bar{\boldsymbol{\phi}}_i \bar{\boldsymbol{\phi}}_i^H \boldsymbol{\Lambda}^{1/2})\mathbf{u}.
\end{align}
The vector $\mathbf{u}$ that maximizes $\mathbf{u}^H \mathbf{A}_m \mathbf{u}$ is the associated eigenvector $\mathbf{q}_M$ of $\lambda_M$. Writing $\mathbf{u}=\sum_{i=1}^M \alpha_i \boldsymbol{\Lambda}^{-1/2} \bar{\boldsymbol{\phi}}_i$ and using the fact that $\bar{\boldsymbol{\phi}}_i$s are orthonormal vectors, we have
\begin{align}
&\mathbf{u}^H \mathbf{A}_m \mathbf{u}=\mathbf{u}^H  (\alpha_m \boldsymbol{\Lambda}^{1/2} \bar{\boldsymbol{\phi}}_m -x \sum_{i\neq m} \alpha_i \boldsymbol{\Lambda}^{1/2}  \bar{\boldsymbol{\phi}}_i )\nonumber \\
&=\alpha_m^2-x\sum_{i\neq m} \alpha_i^2,
\end{align}
The maximum value $\mathbf{u}^H \mathbf{A}_m \mathbf{u}$ takes is $\alpha_m^2$, which happens when $\sum_{i\neq m} \alpha_i^2=0$, i.e. $\alpha_i=0$ for $i\neq m$. As a result $\lambda_M=\alpha_m^2$. The expression for $\alpha_m$ can be obtained by noting that $||\mathbf{u}||_2=1$, so $\alpha_m=\frac{1}{\sqrt{\bar{\boldsymbol{\phi}}_m^H \boldsymbol{\Lambda}^{-1} \bar{\boldsymbol{\phi}}_m}}$. Using this in $\mathbf{u}=\sum_{i=1}^M \alpha_i \boldsymbol{\Lambda}^{-1/2} \bar{\boldsymbol{\phi}}_i$, we can obtain the expression of $\mathbf{q}_M$.

\vspace{-.1in}
\section{Proof of Theorem \ref{Thm2}}
\label{App:Th2}
According to Lemma \ref{Lemmagr}, $\underset{k=1,\dots,K}{\text{max}}\gamma_{k,m}-l_{K,m}$ converges in distribution to a limiting RV.  Solving for $l_{K,m}$ using Lemma \ref{Lemma1} and (\ref{CDF_RBF}), while noting that $\lambda_i$s are functions of $l_{K,m}$ we obtain
\begin{align}
&\frac{\lambda_M(l_{K,m})}{\text{det}(\mathbf{R}) l_{K,m}^{M-1}} \left( \prod_{i=1}^{M-1} \frac{\lambda_i(l_{K,m}) \lambda_M(l_{K,m})}{(\lambda_i(l_{K,m})-\lambda_M(l_{K,m}))}\right) \nonumber \\
&\exp\left(-\frac{l_{K,m}}{\frac{ P_T}{\sigma^2 M} \lambda_M(l_{K,m})}\right)=\frac{1}{ K}.
\end{align}
Manipulation of this expression and defining $c_{l_{K,m}}=\left( \prod_{i=1}^{M-1} \frac{\lambda_i(l_{K,m}) \lambda_M(l_{K,m})}{(\lambda_i(l_{K,m})-\lambda_M(l_{K,m}))}\right)$ yields,
\begin{align}
\label{exp8}
&l_{K,m}=\frac{P_T}{\sigma^2 M} \lambda_M(l_{K,m}) \log K-\frac{P_T}{\sigma^2 M} \lambda_M(l_{K,m})\nonumber \\
&\log\left(\frac{(l_{K,m})^{M-1} \text{det}(\mathbf{R})}{\lambda_M(l_{K,m}) c_{l_{K,m}}} \right).
\end{align}
To this end, note that Lemma \ref{LemmaLamb} implies $\lambda_M(x)$ does not depend on $x$. Utilizing the expression of  $\lambda_M$ from Lemma \ref{LemmaLamb} we obtain
$l_{K,m}=\frac{P_T}{\sigma^2 M} \frac{1}{\bar{\boldsymbol{\phi}}_m^H \boldsymbol{\Lambda}^{-1}\bar{\boldsymbol{\phi}}_m } \log K + O(\log \log K)$, for large $K$ and fixed $M$ and $N$. Substituting $l_{K,m}$  for $\underset{k=1,\dots,K}{\text{max}}\gamma_{k,m}$ in (\ref{exp1}), we obtain the scaling law  as
\begin{align}
\label{expp}
&R_{RBF}^{(K)}=\sum_{m=1}^M \mathbb{E}_{\boldsymbol{\phi}_m}\Big[ \log \Big(1+ \frac{P_T}{\sigma^2 M \bar{\boldsymbol{\phi}}_m^H \boldsymbol{\Lambda}^{-1}\bar{\boldsymbol{\phi}}_m }\log K \nonumber \\
&+ O(\log \log K) \Big)  \Big]+o(1).
\end{align}
Next note that $\bar{\boldsymbol{\phi}}_m^H \boldsymbol{\Lambda}^{-1}\bar{\boldsymbol{\phi}}_m=\boldsymbol{\phi}_m^H \mathbf{R}^{-1}\boldsymbol{\phi}_m$. By defining $\bar{\mathbf{R}}=\frac{1}{\sum_{n=1}^N \alpha_n^2 \beta_r+\beta_d}\mathbf{R}$, we obtain $\bar{\boldsymbol{\phi}}_m^H \boldsymbol{\Lambda}^{-1}\bar{\boldsymbol{\phi}}_m=\frac{1}{ \beta_r\sum_{n=1}^N \alpha_n^2+\beta_d}\boldsymbol{\phi}_m^H \bar{\mathbf{R}}^{-1}\boldsymbol{\phi}_m$. Substituting this back in \eqref{expp} completes the proof.

\section{Proof of Theorem \ref{Cor1}}
\label{App:Lemma2}
Denote the eigenvalue decomposition of $\bar{\mathbf{R}}$ as $\bar{\mathbf{R}}=\mathbf{U}^H \bar{\boldsymbol{\Lambda}} \mathbf{U}$ and define $\bar{\boldsymbol{\phi}}_m=\mathbf{U}\boldsymbol{\phi}_m$. The CDF expression for $y=\frac{1}{\boldsymbol{\phi}_m^H \bar{\mathbf{R}}^{-1}\boldsymbol{\phi}_m}=\frac{1}{\bar{\boldsymbol{\phi}}_m^H \bar{\boldsymbol{\Lambda}}^{-1}\bar{\boldsymbol{\phi}}_m}$ is  given as $F_Y(y)=1-\sum_{i=1}^{M} \eta_i \left(\frac{1}{y}-\frac{1}{\lambda_i(\bar{\boldsymbol{\Lambda}})}  \right)^{M-1}  u\left(1-\frac{y}{\lambda_i(\bar{\boldsymbol{\Lambda}})}  \right)$, where $\eta_i=\frac{1}{\prod_{j \neq i} \left(\frac{1}{\lambda_j(\bar{\boldsymbol{\Lambda}})}-\frac{1}{\lambda_i(\bar{\boldsymbol{\Lambda}})}  \right)}$ and $\lambda_i(\bar{\boldsymbol{\Lambda}})$ is the $i^{th}$ diagonal element of $\bar{\boldsymbol{\Lambda}}$ ordered as $\lambda_1(\bar{\boldsymbol{\Lambda}}) <\dots < \lambda_M(\bar{\boldsymbol{\Lambda}})$. Using this CDF, \normalsize
\begin{align}
&\mathbb{E}\left[\log y\right]=F_Y(y)\log(y)|_{\lambda_1(\bar{\boldsymbol{\Lambda}})}^{\lambda_M(\bar{\boldsymbol{\Lambda}})}-\int_{\lambda_1(\bar{\boldsymbol{\Lambda}})}^{\lambda_M(\bar{\boldsymbol{\Lambda}})} F_Y(y) \frac{1}{y}dy, \nonumber\\
&=\log(\lambda_M(\bar{\boldsymbol{\Lambda}})) -\int_{\lambda_1(\bar{\boldsymbol{\Lambda}})}^{\lambda_M(\bar{\boldsymbol{\Lambda}})} \frac{1}{y} dy + \sum_{i=1}^M \eta_i \int_{\lambda_1(\bar{\boldsymbol{\Lambda}})}^{\lambda_i(\bar{\boldsymbol{\Lambda}})}\frac{1}{y} \nonumber \\
&\left(\frac{1}{y}-\frac{1}{\lambda_i(\bar{\boldsymbol{\Lambda}})}  \right)^{M-1} dy, \nonumber \\
&=\log(\lambda_1(\bar{\boldsymbol{\Lambda}})) +\sum_{i=1}^N \eta_i \int_{\lambda_1(\bar{\boldsymbol{\Lambda}})}^{\lambda_i(\bar{\boldsymbol{\Lambda}})} \sum_{l=0}^{M-1} \binom{M-1}{l} \frac{1}{y^{l+1}}\nonumber \\
&\left( -\frac{1}{\lambda_i(\bar{\boldsymbol{\Lambda}})} \right)^{M-1-l} dy,\nonumber  \\
&=\log(\lambda_1(\bar{\boldsymbol{\Lambda}}))+\sum_{i=1}^N \eta_i  \left( \frac{-1}{\lambda_i(\bar{\boldsymbol{\Lambda}})}  \right)^{M-1} \log y|_{\lambda_1(\bar{\boldsymbol{\Lambda}})}^{\lambda_i(\bar{\boldsymbol{\Lambda}})}\hspace{-.04in}\nonumber \\
&+\sum_{i=1}^N  \eta_i \sum_{l=1}^{M-1}  \binom{M-1}{l} \frac{y^{-l}}{(-l)}\left( \frac{-1}{\lambda_i(\bar{\boldsymbol{\Lambda}})} \right)^{M-1-l}|_{\lambda_1(\bar{\boldsymbol{\Lambda}})}^{\lambda_i(\bar{\boldsymbol{\Lambda}})}, \nonumber 
\end{align} 
\begin{align}
&=\log(\lambda_1(\bar{\boldsymbol{\Lambda}}))+\sum_{i=1}^N \eta_i  \left(-\frac{1}{\lambda_i(\bar{\boldsymbol{\Lambda}})}  \right)^{M-1} \log \left( \frac{\lambda_i(\bar{\boldsymbol{\Lambda}})}{\lambda_1(\bar{\boldsymbol{\Lambda}})} \right) \nonumber \\
&+\sum_{i=1}^N \eta_i \sum_{l=1}^{M-1} \frac{1}{l}\left(\frac{1}{\lambda_1(\bar{\boldsymbol{\Lambda}})^l}-\frac{1}{\lambda_i(\bar{\boldsymbol{\Lambda}})^l}\right)\nonumber \\
& \binom{M-1}{l}\left( -\frac{1}{\lambda_i(\bar{\boldsymbol{\Lambda}})} \right)^{M-1-l}. \nonumber
\end{align}
\bibliographystyle{IEEEtran}
\bibliography{bib}

\begin{IEEEbiography}[{\includegraphics[width=1 in, height=2 in,clip,keepaspectratio ]{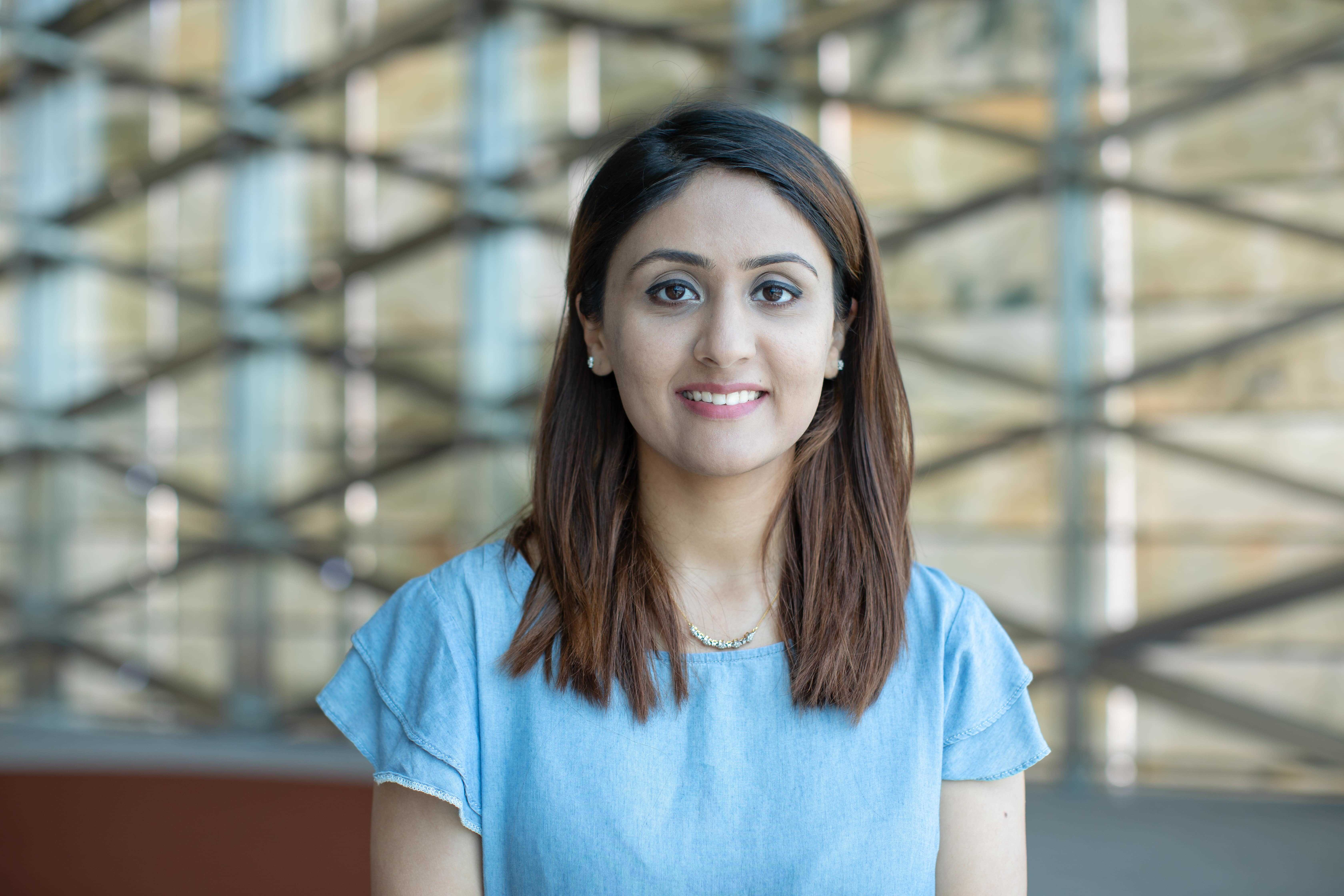}}]{Qurrat-Ul-Ain Nadeem}
(S'15, M'19) was born in Lahore, Pakistan. She received the B.S. degree in electrical engineering from Lahore University of Management Sciences (LUMS), Pakistan in 2013 and the M.S. and Ph.D. degrees in electrical engineering from King Abdullah University of Science and Technology (KAUST), Saudi Arabia in 2015 and 2018 respectively. She is currently a Post-Doctoral Research Fellow with the School of Engineering at the University of British Columbia, Canada. She was selected as the Paul Baran Young Scholar by The Marconi Society in 2018 for her work on full-dimension MIMO. Her research interests include random matrix theory, beamforming design and performance analysis of wireless communication systems.
\end{IEEEbiography}

\begin{IEEEbiography}[{\includegraphics[width=1in,height=1.25in,clip,keepaspectratio]{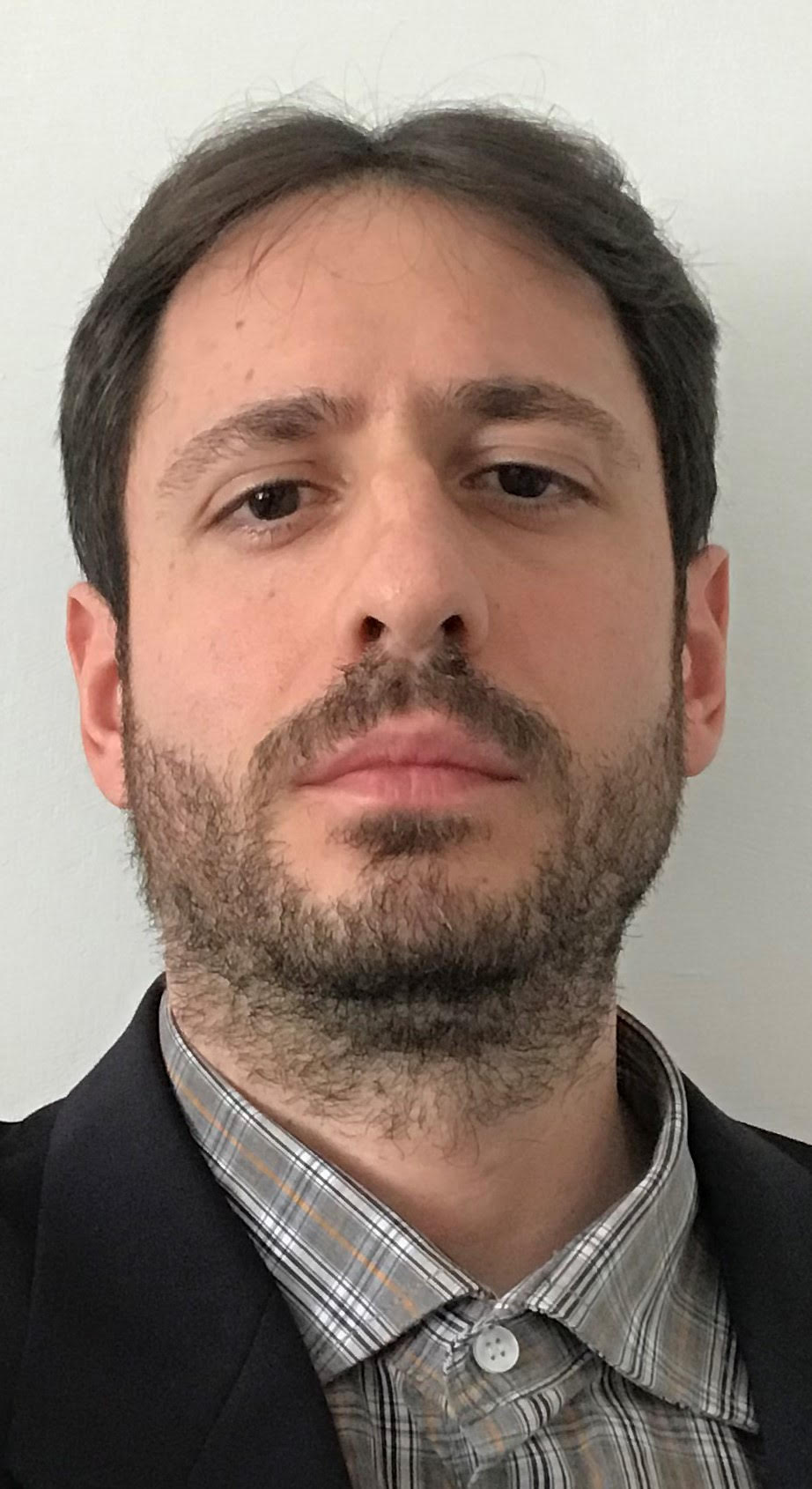}}]{Alessio Zappone} (SM'16) received his M.Sc. and Ph.D. both from the University of Cassino and Southern Lazio (Cassino, Italy). In 2012 he has worked with the Consorzio Nazionale Interuniversitario per le Telecomunicazioni (CNIT) in the framework of the FP7 EU-funded project TREND. From 2012 to 2016 he has been with the Dresden University of Technology, managing the project CEMRIN, funded by the German research foundation (DFG). In 2017 he was the recipient of the H2020 Marie Curie IF BESMART fellowship for experienced researchers, carried out at the LANEAS group of CentraleSupelec (Gif-sur-Yvette, France). Since 2019, he is a tenured professor at the University of Cassino and Southern Lazio.

His research interests lie in the area of communication theory and signal processing, with main focus on optimization techniques for resource allocation and energy efficiency maximization. He held several research appointments at international institutions. He was appointed exemplary reviewer for the \textsc{IEEE Transactions on Communications}  and the \textsc{IEEE Transactions on Wireless Communications} multiple times. He serves as senior area editor for the \textsc{IEEE Signal Processing Letters} and has served as guest editor for the \textsc{IEEE Journal on Selected Areas on Communications} (Special Issues on Energy-Efficient Techniques for 5G Wireless Communication Systems and on Wireless Networks Empowered by Reconfigurable Intelligent Surfaces). He serves as the chair of the special interest group REFLECTIONS on RIS, activated within the SPCE technical committee, and as vice-chair of the IEEE emerging technology initiative on RIS.
\end{IEEEbiography}

\begin{IEEEbiography}[{\includegraphics[width=1 in, height=2 in,clip,keepaspectratio ]{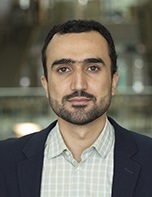}}]{Anas Chaaban} (S'09 - M'14 - SM'17) received the Ma{\^i}trise {\`e}s Sciences degree in electronics from Lebanese University, Lebanon, in 2006, the M.Sc. degree in communications technology and the Dr. Ing. (Ph.D.) degree in electrical engineering and information technology from the University of Ulm and the Ruhr-University of Bochum, Germany, in 2009 and 2013, respectively. From 2008 to 2009, he was with the Daimler AG Research Group On Machine Vision, Ulm, Germany. He was a Research Assistant with the Emmy-Noether Research Group on Wireless Networks located at the University of Ulm, Germany, from 2009 to 2011, and at the Ruhr-University of Bochum from 2011 to 2013. He was a Postdoctoral Researcher with the Ruhr-University of Bochum from 2013 to 2014, and with King Abdullah University of Science and Technology from 2015 to 2017. He joined the School of Engineering at the University of British Columbia as an Assistant Professor in 2018. His research interests are in the areas of information theory and wireless communications.

\end{IEEEbiography}

\end{document}